\documentclass[11pt,a4paper]{article}
\usepackage{color}
\usepackage{amsmath}
\usepackage{amssymb}
\usepackage[utf8]{inputenc}
\usepackage{amsfonts}
\usepackage[header, toc]{appendix}

\newcommand{\be}{\begin{equation}}
\newcommand{\ee}{\end{equation}}
\newcommand{\ba}{\begin{eqnarray}}
\newcommand{\ea}{\end{eqnarray}}

\begin{document}

\begin{center}


{\Large \bf Electric Charge transfer by Four Vector Bosons}\\	
{\large R. Doria\footnote{doria@aprendanet.com} and}
{\large L.S. Mendes \footnote{santiago.petropolis@outlook.com}}\\[0.5cm]

{\large $^1$Aprendanet, Petrópolis, Brazil; Quarks, Petropolis, Brazil}\\
{\large $^2$Aprendanet, Petrópolis, Brazil; CEFET-RJ, Petropolis, Brazil}
\end{center}







\date{\today}

\begin{abstract}

A general theory of electric charge is proposed. It is based on two phenomenologies. Electric charge mutation and conservation law. Three charges $\{ +, - ,0\}$ transformations physics succeeds. Quantum field theory underlies corresponding creations and annihilation. A potential field's quadruplet is ruled. Microscopic electromagnetism is processed by four vectors bosons intermediations. 

The electromagnetism closure is accomplished. The quadruplet $A_{\mu I} \equiv \{ A_\mu, U_\mu, V_\mu^\pm\}$ completeness introduces the most generic EM energy flux between electric charges. Charge mutation includes that besides usual photon, EM phenomena is enlarged by massive and charged photons. Charge conservation associates these four vector fields. Electric charge symmetry, extends EM for an abelian symmetry $U_{Q} \equiv U(1) \times SO(2)_{global}$. A new EM Lagrangian beyond Maxwell results. A symmetry equation for electric charge is established through Noether theorem.

The electric charge transfer physics extends the EM phenomenon. Nonlinear Electromagnetic fields modified electric charge symmetry, new EM regimes. Potential fields become a physical entity producing conglomerates, collective fields, mass, sources, charges, monopoles, forces. EM features ruled from an extended electric charge abelian symmetry. Systemic, nonlinear, neutral, spintronics, photonics, electroweak EM regimes are constituted.
\end{abstract}




\pagestyle{myheadings}


\section{Introduction}

Electromagnetism is the manifestation of electric and magnetic fields. It is based on electric charge and spin. Maxwell equations are its standard description [1]. A qualitative difference appears to Newton’s laws. Their features are that nature is not only constituted by matter, but also through Faraday’s lines of force [2].

Nevertheless, although all its success, Maxwell equations contain limitations. It is a linear theory, introduces the potential fields as a subsidiary, polarization and magnetization fields by hand, passive light, others. Different subjects as condensate matter, plasma, superconductivity, astrophysics are registering such restrictions. There is a physics beyond Maxwell to be encompassed. Something is expected to be done with contemporary electromagnetism. Advances are welcomed since preserving the four Maxwell postulates. Consider light invariance, set EM physics, electric and magnetic fields in pairs, electric charge conservation. There are 52 effective non-Maxwellians extensions, organized into four categories. They are 12 extensions considering the Standard Model, 14 extensions motivated by a physics beyond the Standard Model, 16 nonlinear extensions, and 10 due to dimensionality [3]. The quire is whether the $21^{th}$ century is in front of new electromagnetism or just expecting extensions to Maxwell theory. The effort here will be to enlarge the EM phenomena with a fundamental model. 

Historically, EM development may be divided into three phases. The first one based on charges and magnets was described by William Gilbert in 1600 [4]. The second, in 1865 due to Maxwell, was based on charges and fields [5]. And by third, it is hoped as the $21^{th}$ century development, a relationship between fields and fields. Consider the EM energy flux just depending on fields. A relationship between fields and matter beyond energy and mass. Nonlinear electromagnetism introduces matter dematerialization through EM fields.

The pos-Maxwell concerns to the microscopic understanding of electric charge. It appeared at $20^{th}$ century through the elementary particle physics. Electromagnetism is the theory describing the relationships between electric and magnetic fields. Faraday and Maxwell studied at macroscopic level. The development of elementary particle physics introduced a microscopic electromagnetism. The electric charge being ported by particles with different flavours and spins.

Since 1874 the Irish physicist George Stoney had been suspecting of an electric charge atom, naming it electron [6]. In 1894 stipulated its charge value nearly $ e = 0.5 \cdot 10^{-19} C$. An assignment was realized with the electron discovery by Thomson in 1897 and by Millikan measurement in 1913 of magnitude $e=1.6\cdot10^{-19}C$ in 1913 [7]. The microscopic electromagnetism becomes realistic. Four charges feature followed. Electric charge quantization, carried by particles with different flavors and spins, charges exchanges $\Delta Q = 0, \pm1$, and spin as a magnet.

The $20^{th}$ century did the EM qualitative change from Maxwell macroscopic to a microscopic behavior. The following stages were developed. First, in 1905, Einstein relativity introduced EM fields under reference systems [8]. In 1921 and 1923, Compton proposes the magnetic electron and incorporated the photon as a particle, and in 1924, de Broglie introduced the wave-particle dualism [9]. Third, in 1927, Dirac studies the Maxwell field under second quantization [10]. In 1928, Dirac equation introduces relativistic quantum mechanics for electron particle quantization [11]. And so, at the end of the 1920s and the beginning of the 1930s, the first QED phase was studied. It was a sequence of works with relativistic fields equations by Jordan, Klein, Wigner, Pauli, Heisenberg, Fermi, Fock, Podolski, Weisskopf, Stueckelberg [12]. In the forties, the second QED phase was developed. First, proposed by Schwinger with spin-1 mesotrons (1940) [13] and Dirac with Lorentz group (1942) [14]. Second, after the war, through Tomonaga (1946-1949) [15], Feynman (1948-1951) [16], Schwinger (1948-1954) [17], Dyson (1949) [18]. Third, continued with applications in the 1950s by Landau and others [19-21].

The next stage was scalar electrodynamics by Salam in the fifties [22]. Vector bosons electrodynamics by Salam, Lee, Yang, and others appeared in the sixties [23-29]. Nevertheless, all these quantum fields theories did not surpass the Maxwell boundaries. Although extending Maxwell equations, they were not able to incorporate new EM regimes. For instance, weak interactions as part of electromagnetism. And so, they were replaced by the Standard Model [30-32].

Parallelly, effective nonlinear electromagnetism has been developed since 1925 due to the photon-photon collision. It was introduced by Schapocshnikov, Vavilov, Delbruck, Breit, Wheeler [33]. The $\gamma \gamma -$ scattering motivated nonlinear extensions to Maxwell equations [34]. Diverse scenarios based on effective NLED (nonlinear electrodynamics) were developed since Born-Infeld and Euler-Heisenberg [35]. Actually, there are NLED types as ED ModMax, ED Axionic, ED Logarithmic, and so on [36].

Nevertheless, between these EM periods, a phenomenological missing happened. It was on the electric charge transference. Prospect the electric charge set $\{+ ,0,-\}$ exchanges. Instead of considering the reaction $n \to p + e^- + \tilde{\nu}_e$ through Fermi theory of nuclear beta decay, propose the physics of electric charge transfer. Understand the neutron transmutation into a proton with emission of an electron and neutron as an EM regime with a coupling constant diverse from electric charge. Assuming that, charge transference is an EM phenomenon, and not, a distinct $G_{F}$ weak interaction. 

The phenomenology of electric charge mutation was already a theme for 1923. A time with an electron, proton, photon exchanging three charges. Later in the 1930s, the positron discovery [37], neutron decay [38], muon decay [39], pions cosmic rays decays [40] were confirmed on charges transference. In 1938, by occasion of the Warsaw conference 'New theories in Physics', there were 14 particles in the physics variety. They were these particles $e^-$,$e^+$,$p (p^-)$,$\gamma$,$n(\tilde{n})$,$\mu^-$,$\mu^+$,$\pi^+$,$\pi^0$,$\pi^-$,$\nu(\tilde{\nu})$. There was a rich particle's spectrum confirming the charges set $\{+,0,-\}$ phenomenology.  

At that time, two possibilities to do physics were available. Consider these particles sharing charge mutation, or look for forces of unification. Although different particles reactions coming from cosmic rays and accelerators were showing on charges transference, prevailed the unification scheme. The research line taken by Oscar Klein was unification in five dimensions [41]. The reductionist proposal of unifying four forces becomes the zeitgeist. Klein wicked the name Theory of Everything, ToE [42]. The electric charge mutation argument remained secondary. Although in 1940, Schwinger added mesotrons as spin-1 particles, introducing the existence of two charged vector bosons plus photon, the presence of four-vector bosons intermediating electric charges just appeared in the fifties as belonging to weak interactions.

As time goes by, the view of electric charge mutation was being left. QED took the place for charges exchanges. It was successfully formulated as the theory of electrons and photons. The three charges of electromagnetism were unexplored. Forgotten. The completeness of a whole quantum field theory under charges creation and destruction carried by a set with four intermediate vector particles is still expected. There is a missing EM energy to be considered. An EM energy where each charge is transformed through four-vector bosons $\{A_\mu,U_\mu,V_\mu^\pm \}$. A purpose containing $A_\mu$ as the usual photon, $U_\mu$ massive photon, and $V_\mu^\pm$ massive vector photons. 

Thus, without violating any basic Maxwell principle, an EM enlargement is projected beyond Maxwell and QED. It is called four bosons electromagnetism [43]. While at the macroscopic level, the Coulomb balance did not register on the zero electric charge presence, at a microscopic level, its physicality appears through bosons exchanges. The three electric charge transformations are displayed on the zero charge EM energy. A result already detected at QED with photons carrying a zero charge EM energy. 

This work objective is to investigate a new EM performance under charge transfer. Given the electric charge, mutation to search for new electromagnetic edges. In section 2, one studies the four potential fields association through an extended Abelian gauge symmetry $U_{Q} \equiv U(1) \times SO(2)_{global} $ and the corresponding observable in a Lagrangian. Euler-Lagrangian equations of motion are derived in section 3. Section 4 electric charge physicality is investigated through Noether theorem and identified as the symmetry equation. Section 5 introduces the electromagnetic whole equations of motion. Bianchi's identities are left for section 6. In section 7 invariants are considered. Nonlinear electromagnetic waves are derived in section 8. In section 9, Lorentz's force is extended to the field set. In a conclusion, new perspectives on electromagnetism are considered. Based on nonlinear fields, new electromagnetic regimes, and extended electric charge behavior. 

\section{Four Bosons Lagrangian}

The purpose here is a General Electric charge theory. Consider that electric charge alongside the conservation law contains the mutation factor. Two aspects are supporting such charge exchange physics. The diversity of elementary particles dressing positive, negative, zero charges and the notion of creation and annihilation from quantum field theory. Even so, history shows, the physics of electric charge mutation was not assigned. Physics developed QED just based on positive and negative charges plus the photon. However, despite all the (g-2) QED success [44-46], the existence of an EM completeness intermediate by four bosons is still a missing point to be excavated. 

There is an EM energy to be explored. The electric charge mutation and conservation join together for developing the charges set $\{+,0,-\}$ physics carried by four bosons messengers $\{A_\mu, U_\mu, V_\mu^\pm \}$. A relationship where the gauge principle is extended through the introduction of a family of potential fields in association with an extended Abelian symmetry. An approach has been supported under several points of view [47]. It yields Maxwell being extended by nonlinear Abelian gauge theory. The usual photon field $A_\mu$ appears associated with a massive photon $U_\mu$ and two charged photons $V_\mu^\pm$. 

The electric charge closure is focused. A structure systematized by a quadruplet. While charge mutation requires four intermediate fields, conservation makes their association. Following Noether theorem [48], the quadruplet should be connected through a common gauge parameter. It yields an extended Abelian gauge model $U_{Q} \equiv U(1)\times SO(2)_{global}$ [43], 
\begin{eqnarray}
	A'_\mu &=& A_\mu + k_1 \partial_\mu \alpha \label{Usual photon}\\ 
	U'_\mu &=& U_\mu + k_2 \partial_\mu \alpha \\
	V_\mu^{+'} &=& e^{iq\alpha} \left( V_\mu^{+} +k_{+} \partial_\mu \alpha\right) \\
	V_\mu^{-'} &=& e^{-iq\alpha} \left( V_\mu^- + k_{-} \partial_\mu \alpha\right) \label{Charge photon}
\end{eqnarray}

Eqs. (\ref{Usual photon} - \ref{Charge photon}) introduce the physical fields. They have been  defined as the poles of two points Green's functions. We should remember that different fields' basis was previously constructed in order to obtain the above physical basis [49]. They are consistent with Kamechu-Raifeartaigh-Salam prescriptions [50]. Consequently, $k_1,k_2, k_\pm$ are terms related to parameters at the field basis transformations matrix. They are depending on theory-free coefficients associated with Lagrangian terms [51].

EM physics is extended to a quadruplet $A_{\mu I} \equiv \{ A_{\mu}, U_{\mu}, V_{\mu}^{\pm}\}$. Interconnected potential fields share an Abelian symmetry. Observable is formed. Fields conglomerates, granular and collective fields strengths are derived. Potential fields conglomerates are fields linear combinations, as $\mathbf{e}_{[IJ]}A_\mu^J$,  where $\mathbf{e}_{[IJ]}$ is a parameter on free coefficients. Granular and collective fields will be described below. Their corresponding gauge invariance properties are studied in Appendix A.

For the antisymmetric sector, one gets the following granular fields,

\begin{eqnarray}
	F_{\mu \nu} &=   \partial_\mu A_\nu - \partial_{\nu}A_\mu , & F'_{\mu \nu} = F_{\mu \nu} \label{Campo granular primeiro}\\
	U_{\mu \nu} &=  \partial_\mu U_\nu - \partial_{\nu}U_\mu , & U'_{\mu \nu} = U_{\mu \nu}\\
	V_{\mu \nu}^\pm &=  \partial_\mu V_\nu^\pm - \partial_{\nu}V_\mu^\pm, &   V_{\mu \nu}^{\pm'} = e^{\pm i q \alpha } V_{\mu \nu}^\pm \label{Campo granular ultimo}
\end{eqnarray}

The collective antisymmetric field strengths are

\begin{eqnarray}
	&&\mathbf{e}^{[12]}_{[\mu \nu]} = \mathbf{e}_{[12]}\left(A_\mu U_\nu - A_\nu U_\mu \right)
	\\ 
	&&\mathbf{e}^{[+-]}_{[\mu \nu]} = -i\mathbf{e}_{[34]}\left(V_\mu^+ V_\nu^- - V_\nu^+ V_\mu^- \right) 
	\\  
	&&\mathbf{e}^{[+1]}_{[\mu \nu]} +\mathbf{e}^{[-1]}_{[\mu \nu]} = \frac{1}{\sqrt{2}}\big[\left(\mathbf{e}_{[13]} + i \mathbf{e}_{[14]}\right) \left( A_\mu V_\nu^+-A_\nu V_\mu^+ \right)\nonumber
	\\
	&&+  \left(\mathbf{e}_{[13]} - i \mathbf{e}_{[14]}\right) \left( A_\mu V_\nu^--A_\nu V_\mu^- \right)\big]
	 \\ 
	&&\mathbf{e}^{[+2]}_{[\mu \nu]} +\mathbf{e}^{[-2]}_{[\mu \nu]} = \frac{1}{\sqrt{2}}\big[\left(\mathbf{e}_{[23]} + i \mathbf{e}_{[24]}\right) \left( U_\mu V_\nu^+-U_\nu V_\mu^+ \right) + \nonumber
	\\
	 &&\left(\mathbf{e}_{[23]} - i \mathbf{e}_{[24]}\right) \left( U_\mu V_\nu^--U_\nu V_\mu^- \right)\big] 
\end{eqnarray}

For the symmetric sector, the granular fields strengths are
\begin{eqnarray}
	S_{\mu \nu 1} &= \partial_\mu A_\nu + \partial_\nu A_\mu, &  S_{\mu \nu 1}'= S_{\mu \nu 1}\\
	S_{\mu \nu 2} &= \partial_\mu U_\nu + \partial_\nu U_\mu, &  S_{\mu \nu 2}'= S_{\mu \nu 2}\\
	S_{\mu \nu}^{\pm} &=  \partial_\mu V_\nu^{\pm} + \partial_\nu V_\mu^{\pm}, &  S_{\mu \nu }^{\pm'}=e^{\pm i q \alpha} S^{\pm}_{\mu \nu }
\end{eqnarray}
and
\begin{eqnarray}
	S_{\alpha 1 }^{\alpha} &= 2\partial_\alpha A^{\alpha} , &  S_{\alpha 1 }^{\alpha '} =  S_{\alpha 1 }^{\alpha}\\
	S_{\alpha 2 }^{\alpha} &=  \partial_\alpha U^{\alpha}, &  S_{\alpha 2}^{\alpha'}= S_{\alpha 2 }^{\alpha }\\
	S_{\alpha}^{\alpha \pm} &=  \partial_\alpha V^{\alpha \pm}, &  S_{\alpha }^{\alpha \pm'}=e^{\pm i q \alpha} S_{\alpha}^{\alpha \pm}
\end{eqnarray}

The collective symmetric fields strengths are
\begin{eqnarray}
	&&\mathbf{e}^{(11)}_{(\mu \nu)} = \mathbf{e}_{(11)}A_\mu A_\nu 
	\\
	&&\mathbf{e}^{(22)}_{(\mu \nu)} = \mathbf{e}_{(22)}U_\mu U_\nu 
	\\
	&&\mathbf{e}^{(12)}_{(\mu \nu )} = \mathbf{e}_{(12)}\left(A_\mu U_\nu + A_\nu U_\mu \right)
	\\
	&&\mathbf{e}^{(+-)}_{(\mu \nu)} = \frac{1}{2}\left( \mathbf{e}_{(33)} + \mathbf{e}_{(44)}\right)\left( V_\mu^+V_\nu^- + V_\nu^+V_\mu^- \right)
	\\
	&&\mathbf{e}^{(+1)}_{(\mu \nu)}+ \mathbf{e}^{(-1)}_{(\mu \nu)} = \frac{1}{\sqrt{2}}\big[\left(\mathbf{e}_{(13)} + i\mathbf{e}_{(14)}\right)\left(A_\mu V_\nu^+ + A_\nu V_\mu^+ \right) +\nonumber
	\\
	&&+\left(\mathbf{e}_{(13)} - i\mathbf{e}_{(14)}\right)\left(A_\mu V_\nu^- + A_\nu V_\mu^- \right)\big]
	\\
	&&\mathbf{e}^{(+2)}_{(\mu \nu)}+ \mathbf{e}^{(-2)}_{(\mu \nu)} = \frac{1}{\sqrt{2}}\big[\left(\mathbf{e}_{(23)} + i\mathbf{e}_{(24)}\right)\left(U_\mu V_\nu^+ + U_\nu V_\mu^+ \right) +\nonumber
	\\
	&& \left(\mathbf{e}_{(23)} - i\mathbf{e}_{(24)}\right)\left(U_\mu V_\nu^- + U_\nu V_\mu^- \right)\big]
	\\
    &&\mathbf{e}^{(++)}_{(\mu \nu)} + \mathbf{e}^{(--)}_{(\mu \nu)} = \frac{1}{2}\left(\mathbf{e}_{(33)}-\mathbf{e}_{(44)}\right)\left(V_\mu^+V_\nu^+ + V_\mu^-V_\nu^-\right)
	\\
	&&\mathbf{e}^{(++34)(\mu \nu)} + \mathbf{e}^{(--34)(\mu \nu)} =  i\mathbf{e}_{(34)}V_\mu^+V_\nu^+ -i\mathbf{e}_{(34)}V_\mu^-V_\nu^-\label{Collective +-}
\end{eqnarray}
Under the gauge transformations cited in Appendix A. Similarly for scalar symmetric collective terms, eqs. (\ref{Scalar phontos AA}-\ref{Scalar phontos V+V-}) .

The field strengths are written at eqs. (\ref{Campo granular primeiro} - \ref{Collective +-}) are real observable which gauge invariance is studied in Appendix A. Notice that $ \mathbf{e}_{[12]}, \mathbf{e}_{[13]},\mathbf{e}_{(11)}$ and so on are just parameters. They are depending on the free coefficients generated by the model. Consequently, they can take any value without violating the gauge symmetry. This property will support the correspondent gauge invariance.

Thus, an Abelian formulation joining distinct granular and collective fields strengths is obtained. While non-abelian models architect them together as $F_{\mu \nu} = \partial_{\mu}A_{\nu} - \partial_{\nu}A_{\mu} + g[A_\mu , A_\nu]$, eqs. (\ref{Usual photon}-\ref{Charge photon}) split in two separated gauge invariance terms. Two distinct physicality are derived to be understood dynamically.

A new EM energy appears. The following Lagrangian is organized in terms of the above observable and free parameters
\begin{equation}
	L=L_A+L_S+L_M+ L_{GF} \label{Lagrangiana}
\end{equation}

The antisymmetric sector is
\begin{equation}
	L_A=L_A^K+L_A^3+L_A^4
\end{equation}
where
\begin{equation}
	L_A^K = a_1 F_{\mu\nu} F^{\mu\nu} +a_2 U_{\mu\nu}U^{\mu \nu}+a_3 V_{\mu\nu}^+ V^{\mu\nu-}	
\end{equation}
and
\begin{eqnarray}	
	&&L_A^3= \left( b_1 F_{\mu\nu} + b_2 U_{\mu\nu} \right) \left( \mathbf{e}^{[12][\mu\nu]} +\mathbf{e}^{[+-][\mu\nu]}  \right) +\nonumber
	\\
	&&+ b_3 V_{\mu\nu}^+ \left( \mathbf{e}^{[-1][\mu\nu]} +e^{[-2][\mu\nu]}  \right)+ b_3 V_{\mu\nu}^- \left(\mathbf{e}^{[+1][\mu\nu]} +\mathbf{e}^{[+2][\mu\nu]} \right)
\end{eqnarray}
and
\begin{equation}
	L_A^4=\left( \mathbf{e}^{[12][\mu\nu]} +\mathbf{e}^{[+-][\mu\nu]}  \right)^2 + \left( \mathbf{e}^{[-1][\mu\nu]} +e^{[-2][\mu\nu]}  \right)\left(\mathbf{e}^{[+1][\mu\nu]} +\mathbf{e}^{[+2][\mu\nu]} \right)
\end{equation}	

The symmetric sector is
\begin{equation}
	L_S=L_S^K+L_S^3+L_S^4
\end{equation}
where
\begin{eqnarray}
	&&L_S^K = \beta_1 S_{\mu\nu1}\ S^{\mu\nu1}+\beta_2 S_{\mu\nu2} S^{\mu\nu2}+\beta_+ \beta_- S_{\mu\nu}^+ S^{\mu\nu-} + \nonumber
	\\
	&&+\big( \rho_1 g^{\mu\nu} S_\alpha^{\alpha1}+\rho_2 g^{\mu\nu} S_\alpha^\alpha\big)^2+16\rho_+ \rho_- S_{\alpha}^{\alpha +} S_{\beta}^{\beta-}  + \rho_+ \beta_- g^{\mu\nu} S_\alpha^{\alpha+} S_{\mu\nu}^-
	\\
	&& \left( \rho_1 g^{\mu\nu} S_{\alpha1}^\alpha+\rho_2 g^{\mu\nu} S_{\alpha2}^\alpha \right) \left( \beta_1 S_{\mu\nu 1}+\beta_2 S_{\mu\nu2}\right)+ \rho_- \beta_+ g^{\mu\nu} S_\alpha^{\alpha -} S_{\mu\nu}^+\nonumber
\end{eqnarray}
and
\begin{eqnarray}
	&&L_S^3=\left(\beta_1 S_{\mu\nu 1}+\beta_2 S_{\mu\nu 2}+ \rho_1 g^{\mu\nu} S_{\alpha1}^\alpha +\rho_2 g^\mu\nu S_{\alpha2}^\alpha \right)\big( \mathbf{e}^{(11)(\mu\nu)} 
	\nonumber
	\\
	&&++\mathbf{e}^{(22)(\mu\nu)} +g^{\mu\nu} \mathbf{e}^{(11)\alpha}_\alpha +g^{\mu\nu} \mathbf{e}^{(22)\alpha}_\alpha + \mathbf{e}^{(12)(\mu\nu)} +g^{\mu\nu} \mathbf{e}^{(12)\alpha}_\alpha  \nonumber
	\\
	&&+ \mathbf{e}^{(+-)(\mu\nu)} +g^{\mu\nu} \mathbf{e}^{(+-)\alpha}_{\alpha} \big)+\left(\beta_+ S_{\mu\nu}^+ +\rho_+ g^{\mu\nu} S_\alpha^{\alpha+} \right) \big(\mathbf{e}^{(-1)(\mu\nu)} 
	\\
	&&+g^{\mu\nu} \mathbf{e}^{(-1)\alpha}_\alpha + \mathbf{e}^{(-2)(\mu\nu)} + g^{\mu\nu} \mathbf{e}^{(-2)\alpha}_\alpha \big) + \left(\beta_- S_{\mu\nu}^- + \rho_- g^{\mu\nu} S_\alpha^{\alpha- } \right)\nonumber
	\\
	&&\big(\mathbf{e}^{(+1)(\mu\nu)} +g^{\mu\nu} \mathbf{e}^{(+1)\alpha}_{\alpha} + \mathbf{e}^{(+2)(\mu\nu)} +g^{\mu\nu} \mathbf{e}^{(+2)\alpha}_\alpha \big)\nonumber
\end{eqnarray}
and
\begin{eqnarray}
	&&L_S^4= \big( \mathbf{e}^{(11)(\mu\nu)} +g^{\mu\nu} \mathbf{e}^{(11)\alpha}_\alpha
	+\mathbf{e}^{(22)(\mu\nu)} +g^{\mu\nu} \mathbf{e}^{(22)\alpha}_\alpha +
	\mathbf{e}^{(12)(\mu\nu)} +\nonumber
	\\
	&& +g^{\mu\nu} \mathbf{e}^{(12)\alpha}_\alpha + \mathbf{e}^{(+-)(\mu\nu)}+g^{\mu\nu} \mathbf{e}^{(12)\alpha}_\alpha + \mathbf{e}^{(+-)(\mu\nu)}+g^{\mu\nu} \mathbf{e}^{(+-)\alpha}_{\alpha} \big)^2+\nonumber
	\\
	&&\left(\mathbf{e}^{(-1)(\mu\nu)} +g^{\mu\nu} \mathbf{e}^{(-1)\alpha}_\alpha + \mathbf{e}^{(-2)(\mu\nu)} + g^{\mu\nu} \mathbf{e}^{(-2)\alpha}_\alpha \right)\big(\mathbf{e}^{(+1)(\mu\nu)} 
	\\
	&&+g^{\mu\nu} \mathbf{e}^{(+1)\alpha}_{\alpha}+\mathbf{e}^{(+2)(\mu\nu)} +g^{\mu\nu} \mathbf{e}^{(+2)\alpha}_\alpha \big)+ \mathbf{e}_{(\mu \nu)}^{(++)} \mathbf{e}^{(--)(\mu \nu)} +\nonumber
	\\
	&&+ 16 \mathbf{e}_{(\mu \nu)}^{(++)} \mathbf{e}^{(--)(\mu \nu)} + \mathbf{e}_{(\mu \nu)}^{(++34)} e^{(--34)(\mu \nu)}+ 16\mathbf{e}_{(\mu \nu)}^{(++34)} \mathbf{e}^{(--34)(\mu \nu)}\nonumber
\end{eqnarray}

$L^{3}$ and $L^{4}$ develop an Abelian nonlinear EM. A Maxwell extension with a new perspective for Faraday lines of force. Stipulated by electric charge phenomenology, Lagrangian eq. (\ref{Campo granular primeiro}) penetrates in the third EM historical stage. A self-interacting structure was diverse from Yang-Mills.

The mass sector is
\begin{equation}
	L_M=\mathbf{m}^2_U U_\mu U^\mu+\mathbf{m}^2_V V_\mu^+ V^{\mu-} \label{Termo Massa}
\end{equation}
where eq. (\ref{Termo Massa}) introduces a mass term without requiring the spontaneously breaking symmetry [52]. Notice that $\mathbf{m}^2_{U}$ and $\mathbf{m}^2_{V}$ are just mass parameters.

The gauge fixing term is [49]

\begin{eqnarray}
	&&L_{GF}= \frac{1}{4} \xi_{(11)} S_{\alpha1}^\alpha S_{\beta1}^\beta + \frac{1}{4} \xi_{(22)} S_{\alpha2}^\alpha S_{\beta 2}^\beta+ \frac{1}{2} \xi_{(12)} S_{\alpha 1}^\alpha S_{\beta 2}^\beta \nonumber
	\\
	&&+\frac{1}{4} \left( \xi_{(33)} + \xi_{(44)}\right)S_\alpha^{\alpha +}S_{\beta}^{\beta -}
\end{eqnarray}
Notice that there is only one gauge fixing relating the four potential fields. It is due to the presence of just one gauge parameter. However, it provides five arbitrary parameters $\xi_{(11)}, \xi_{(12)}, \xi_{(22)}, \xi_{(33)}, \xi_{(44)} $ which can assume any value. This means that, differently from the usual case, more than one constraint relationship can be taken. 

The four bosons Lagrangian generalizes previous results. In 1962, Lee and Yang introduced the longitudinal bosons propagation and the term $F_{\mu \nu}W^{\mu}W^{\nu*}$ in order to correct the dipole moment magnet for charged bosons [25]. In 1963, Salam introduced the term $\lambda\left( W_\mu W^{\mu *}\right)^2$ to ensure renormalization by power counting [23-24]. These terms are included in $L_A^3$ plus other magnetic moment terms, and $L_S^4$, respectively. Self-interacting photons are included at $L_S^3$ and $L_S^4$. Yet, differently from Maxwell, the model contains spin-1 and spin-0 sectors. There is a more generic EM to be explored.

\section{Euler-Lagrange equations}

Physics requires equations. While in the past, homothety introduced symmetry, modern physics does with symmetry [53]. Trigonometry generated functions like sin and cos, eqs. (\ref{Usual photon}-\ref{Charge photon}) introduces observable. Both provide variables to be analyzed through equations. Trigonometry establish $sin^2(x) + cos^2(x) = 1$. The corresponding minimal action principle produces Euler-Lagrange equations. Consequently, similarly to trigonometry, a mathematical structure based on fields set $\{A_\mu, U_\mu, V_\mu^\pm \}$ is generated. A nonlinear system of hyperbolic differential equations is developed. The following four whole relativistic equations are derived.

For photon-$A_\mu$, 
\begin{eqnarray}
	&&\partial_\nu \big\{4a_1 F^{\nu\mu} + 4\beta_1\ S^{\nu\mu 1}+ 4 \big(\rho_1\ \beta_1+11\rho_1 + \frac{1}{2} \xi_{(11)}  \big) g^{\nu\mu} S_{\alpha 1}^{\alpha} \nonumber
	\\
	&&+2\big(\rho_1 \beta_2+\rho_2 \beta_2+22\rho_1 \rho_2 + \frac{1}{2}\xi_{(12)} \big) g^{\nu \mu} S_{\alpha2}^{\alpha }+2b_1 \big(\mathbf{e}^{[12][\nu \mu]}  \nonumber
	\\
	&&+ \mathbf{e}^{[+-][\nu \mu]}  \big)+34\rho_1 g^{\nu \mu} \big(\mathbf{e}^{(11)\alpha}_\alpha + \mathbf{e}^{(22)\alpha}_\alpha+ \mathbf{e}^{(12)\alpha}_\alpha +\mathbf{e}^{(+-)\alpha}_\alpha \big)\nonumber
	\\
	&&+2\beta_1 \big(\mathbf{e}^{(11)(\nu \mu)}+g^{\nu \mu} \mathbf{e}^{(11)\alpha}_\alpha +g^{\nu \mu} \mathbf{e}^{(22)\alpha}_\alpha+\mathbf{e}^{(12)(\nu \mu)}   
	\\
	&&+g^{\nu \mu} \mathbf{e}^{(12)\alpha}_\alpha +g^{\nu \mu} \mathbf{e}^{(+-)\alpha}_\alpha  +\mathbf{e}^{(+-)(\nu \mu)}\big) \big\}= J_A^\mu \nonumber
\end{eqnarray}
where
\begin{eqnarray}
	&&J_A^\mu = 2\mathbf{e}_{[12]} \left( b_1 F^{\mu \nu}+ b_2U^{\mu \nu} \right)  U_\nu + \sqrt{2} b_3 \left( \mathbf{e}_{[13]} -i\mathbf{e}_{[14]}  \right)  V^{\mu \nu+}  V_\nu^- +\sqrt{2} b_3 \big( \mathbf{e}_{[13]} \nonumber
	\\
	&&+i\mathbf{e}_{[14]}  \big)  V^{\mu \nu -}V_\nu^+ +2\mathbf{e}_{(11)}\left(\beta_1 S_{\mu \nu 1}+\beta_2 S_{\mu \nu 2}+ \rho_1 g^{\mu \nu}  S_{\alpha1}^\alpha+\rho_2 g^{\mu \nu} S_{\alpha 2}^\alpha \right) A_\nu+ 2\mathbf{e}_{(12)}\big( \beta_1 S_{\mu \nu 1}\nonumber
	\\
	&&+\beta_2 S_{\mu \nu 2}+ \rho_1 g^{\mu \nu}  S_{\alpha1}^\alpha+\rho_2 g^{\mu \nu} S_{\alpha 2}^\alpha\big)U_\nu+
	2e_{(12)} \left[\left(\beta_1+16\rho_1 \right) S_{\alpha1}^\alpha+\left(\beta_2+
	16\rho_2 \right) S_{\alpha 2}^{\alpha }\right] U^{\mu}\nonumber  
	\\
	&&	+ \sqrt{2}\left(\mathbf{e}_{13}-i\mathbf{e}_{(14)}\right)\big[ \beta_+S^{\mu \nu+}+\big(16\beta_+ + 17\rho_+\big) g^{\mu \nu}S_\alpha^{\alpha+}\big]V^-_\nu +
	+\sqrt{2}\left(\mathbf{e}_{(13)}+i\mathbf{e}_{(14)}\right)\nonumber
	\\
	&&[ \beta_-S^{\mu \nu-}+ (16\beta_- + 17\rho_-) g^{\mu \nu}S_\alpha^{\alpha-}]V^+_\nu 
	+4\mathbf{e}_{[12]}\left( \mathbf{e}^{[12][\mu\nu]} + \mathbf{e}^{[+-][\mu\nu]} \right)U_\nu+\sqrt{2}\big(\mathbf{e}_{[13]}\nonumber
	\\
	&&-i\mathbf{e}_{[14]}\big)\left(\mathbf{e}^{[+1][\mu\nu]}+  \mathbf{e}^{[+2][\mu\nu]}\right)V_\nu^-
	+\sqrt{2}\big(\mathbf{e}_{[13]}+i\mathbf{e}_{[14]}\big)\left(\mathbf{e}^{[-1][\mu\nu]} +\mathbf{e}^{[-2][\mu\nu]}\right)V_\nu^+ \nonumber
	\\
	&&+4\mathbf{e}_{(11)}(\mathbf{e}^{(11)(\mu \nu)}+ 2g^{\mu \nu} \mathbf{e}^{(11)\alpha}_\alpha+ \mathbf{e}^{(22)(\mu \nu)}+ \mathbf{e}^{(+-)(\mu \nu)}+18g^{\mu \nu}\mathbf{e}^{(22)\alpha}_\alpha +
	\mathbf{e}^{(22)(\mu \nu)} + 
	\\
	&&18g^{\mu \nu}\mathbf{e}^{(12)\alpha}_{\alpha}+17g^{\mu \nu}\mathbf{e}^{(+-)\alpha}_{\alpha})A_\nu + 4\mathbf{e}_{(12)}(\mathbf{e}^{(11)(\mu \nu)} + g^{\mu \nu}\mathbf{e}^{(12)\alpha}_{\alpha} + g^{\mu \nu}\mathbf{e}^{(11)\alpha}_{\alpha}+
	18\mathbf{e}^{(22)(\mu \nu)})U_\nu\nonumber
	\\
	&&\sqrt{2}\left(\mathbf{e}_{(13)} + i\mathbf{e}_{(14)}\right)(\mathbf{e}^{(-1)(\mu \nu)}+18g^{\mu\nu}\mathbf{e}^{(-1)\alpha}_{\alpha})V_\nu^++\sqrt{2}\left(\mathbf{e}_{(13)} - i\mathbf{e}_{(14)}\right)(\mathbf{e}^{(+1)(\mu \nu)} \nonumber
	\\
	&&+18g^{\mu\nu}\mathbf{e}^{(+1)\alpha}_{\alpha})V_\nu^- 
	+ \sqrt{2}\left(\mathbf{e}_{(23)} + i\mathbf{e}_{(24)}\right)(\mathbf{e}^{(-2)(\mu \nu)}+18g^{\mu \nu}\mathbf{e}^{(-2)\alpha}_\alpha )V_\nu^+\nonumber
	\\
	&& 
	+\sqrt{2}\left(\mathbf{e}_{(23)} - i\mathbf{e}_{(24)}\right)(\mathbf{e}^{(+2)(\mu \nu)}
	+18g^{\mu \nu} \mathbf{e}^{(+2)\alpha}_\alpha )V_\nu^- \nonumber 
\end{eqnarray}

For massive photon-$U_\mu$, 

\begin{eqnarray}\label{Euler-Lagrange U}
	&&\partial_\nu \{ 4 a_2 U^{\nu \mu}+ 4\beta_2 S^{\nu\mu 2} + 4\big( \rho_2\beta_2+11\rho_2 + 1/2\xi_{(22)}\big) S_\alpha^{\alpha 2}+\nonumber
	\\
	&&+ 2\big(\rho_2\beta_1+\rho_1\beta_2+22\rho_1\rho_2 + 1/2\xi_{(12)}\big) S_\alpha^{\alpha1}+ 2 b_2 \big(\mathbf{e}^{[12][\nu \mu]}\nonumber
	\\
	&&+\mathbf{e}^{[+-][\nu\mu]}\big)+ 34\rho_2+ g^{\nu\mu}\big(\mathbf{e}^{(11)\alpha}_\alpha+\mathbf{e}^{(22)\alpha}_{\alpha}+\mathbf{e}^{(12)\alpha}_\alpha+\mathbf{e}^{(+-)\alpha}_{\alpha} \big) 
	\\
	&&2\beta_2(\mathbf{e}^{(11)(\nu\mu)}+g^{\nu\mu}\mathbf{e}^{(11)\alpha}_\alpha  +\mathbf{e}^{(22)(\nu\mu)}
	+g^{\nu\mu}\mathbf{e}^{(22)\alpha}_\alpha + \mathbf{e}^{(12)(\nu\mu)}\nonumber 
	\\
	&&+g^{\nu\mu}\mathbf{e}^{(12)\alpha}_\alpha + \mathbf{e}^{(+-)(\nu\mu)}+g^{\nu\mu}\mathbf{e}^{(+-)\alpha}) \} - 2\mathbf{m}_U U^\mu = J_U^\mu \nonumber 
\end{eqnarray}
where
\begin{eqnarray}
	&&J_{U}^\mu=2\mathbf{e}_{[12]}\left(b_1F^{\mu\nu} + b_2 U^{\mu\nu}\right)A_\nu+\sqrt{2}b_3 \left(\mathbf{e}_{[23]}-i\mathbf{e}_{[24]}\right)V^{\mu\nu+}V_\nu^-\nonumber
	\\
	&& +\sqrt{2}b_3 \big(\mathbf{e}_{[23]}+i\mathbf{e}_{[24]}\big)V^{\mu\nu-}V_\nu^+
	+2\mathbf{e}_{(22)}\big[(\beta_2+17\rho_2) g^{\mu \nu}S^\alpha_{\alpha 2}+\nonumber 
	\\
	&& +(\beta_1+17\rho_1)g^{\mu \nu}S^\alpha_{\alpha 1} +\beta_1 S^{\mu \nu}_1 +\beta_2 S^{\mu \nu}_2  \big] U_\nu + 2\mathbf{e}_{(12)}[\beta_1S^{\mu \nu}_1 \nonumber
	\\
	&&+ \beta_2S^{\mu \nu}_2 +(\beta_1+17\rho_1)g^{\mu \nu }S^{\alpha}_{\alpha 1}(\beta_2+17\rho_2)g^{\mu \nu}S^{\alpha}_{\alpha 2}]A_\nu + \nonumber
	\\
	&&+4\mathbf{e}_{[12]}\mathbf{e}^{[12][\mu\nu]}A_\nu+
	\sqrt{2}\left(\mathbf{e}_{(23)} -i\mathbf{e}_{(24)}\right)[\beta_+S^{\alpha +}_\alpha  +\nonumber
	\\
	&&+ (\beta_+ + 17\rho_+)g^{\mu \nu}S^{\alpha +}_{\alpha}]V_\nu^-+\sqrt{2}\left(\mathbf{e}_{(23)} +i\mathbf{e}_{(24)}\right)[\beta_-S^{\alpha -}_\alpha \nonumber
	\\
	&&+ (\beta_-+17\rho_-)g^{\mu \nu}S^{\alpha -}_{\alpha}]V_\nu^+
	+4\mathbf{e}_{[12]}\mathbf{e}^{[+-][\mu\nu]} A_\nu +\nonumber
	\\
	&&\sqrt{2}\left(\mathbf{e}_{[23]}-i\mathbf{e}_{[24]}\right)(\mathbf{e}^{[+1][\mu\nu]} 
	+\mathbf{e}^{[+2][\mu\nu]})V_\nu^-+\nonumber
	\\
	&&\sqrt{2}\left(\mathbf{e}_{[23]}+i\mathbf{e}_{[24]}\right)(\mathbf{e}^{[-1][\mu\nu]} + \mathbf{e}^{[-2][\mu\nu]})V_\nu^+\nonumber
	\\
	&&+ 4\mathbf{e}_{(22)}(2\mathbf{e}^{(22)(\mu \nu)} + 2g^{\mu \nu}\mathbf{e}^{(22)\alpha}_\alpha +
	\mathbf{e}^{(11)(\mu \nu)} + \mathbf{e}^{(+-)(\mu \nu)}
	\\
	&&+ 18g^{\mu \nu}\mathbf{e}^{(11)\alpha}_\alpha + 18 g^{\mu \nu}\mathbf{e}^{(12)\alpha}_\alpha+ 17g^{\mu \nu} \mathbf{e}^{(+-)\alpha}_\alpha) U_\nu\nonumber 
	\\
	&& + 4\mathbf{e}_{(12)}(+18g^{\mu \nu}\mathbf{e}^{(11)\alpha}_\alpha +
	18g^{\mu \nu}\mathbf{e}^{(+-)\alpha}_\alpha +\mathbf{e}^{(22)(\mu \nu)} + g^{\mu \nu}\mathbf{e}^{(12)\alpha}_\alpha \nonumber
	\\
	&&+ 2\mathbf{e}^{(11)(\mu \nu)})A_{\nu} + \sqrt{2}\left(\mathbf{e}_{(23)}+i\mathbf{e}_{(24)} \right)(\mathbf{e}^{(-1)(\mu \nu)}\nonumber
	\\
	&& + 18g^{\mu \nu}\mathbf{e}^{(-1)\alpha}_\alpha) V_\nu^+ +\sqrt{2}\left(\mathbf{e}_{(23)} - i\mathbf{e}_{(24)} \right)(\mathbf{e}^{(+1)} + 18g^{\mu \nu}\mathbf{e}^{(+1)\alpha}_\alpha)V_\nu^- +\nonumber 
	\\
	&&+	\sqrt{2}\left(\mathbf{e}_{(23)} - i\mathbf{e}_{(24)} \right)(\mathbf{e}^{(+2)(\mu \nu)} +18g^{\mu \nu}\mathbf{e}^{(+2)\alpha}_\alpha)V_\nu^- +\nonumber
	\\
	&&+18\sqrt{2}\left(\mathbf{e}_{(23)} +i\mathbf{e}_{(24)} \right)(\mathbf{e}^{(-2)(\mu \nu)} + g^{\mu \nu}\mathbf{e}^{(-2)\alpha}_\alpha) V_\nu^+\nonumber
\end{eqnarray}

For charged photon-fields,
\begin{eqnarray}
	&&\partial_\nu \{2a_3V^{\mu\nu-}+2\beta_+\beta_-S^{\nu\mu-}+2\big(16\rho_+\rho_-+\rho_+\beta_-+\rho_-\beta_+\nonumber 
	\\
	&&1+1/4\left(\xi_{33}+\xi_{44}\right) \big)g^{\nu\mu}S_\alpha^{\alpha-}+2b_3\left(\mathbf{e}^{[-1][\nu\mu]}+\mathbf{e}^{[-2][\nu\mu]}\right)+\nonumber
	\\
	&&+\beta_+\left(\mathbf{e}^{(-1)(\nu\mu)}+g^{\nu\mu}\mathbf{e}^{(-1)\alpha}_\alpha+\mathbf{e}^{(-2)(\nu\mu)}+g^{\nu\mu}\mathbf{e}^{(-2)\alpha}_\alpha \right)+
	\\
	&&+34\rho_+g^{\nu\mu}\left(\mathbf{e}^{(-1)\alpha}_\alpha+ \mathbf{e}^{(-2)\alpha}_\alpha\right)\}-\mathbf{m}_V V^\mu=J_V^{\mu-}\nonumber
\end{eqnarray}
where
\begin{eqnarray}
	&&J_V^{\mu-}=-2i\mathbf{e}_{[34]}\left(b_1F^{\mu\nu}+b_2U^{\mu\nu}\right)V_\nu^-+\sqrt{2}\left(\mathbf{e}_{[13]}+i\mathbf{e}_{[14]}\right)V^{\mu\nu-}A_\nu +\big(\mathbf{e}_{(33)}+\nonumber
	\\
	&&+ \mathbf{e}_{(44)}\big)[\beta_1S^{\mu \nu}_1 +\beta_2S^{\mu \nu}_2 +
	(\beta_1+17\rho_1)g^{\mu \nu}S^{\alpha}_{\alpha 1}
	(\beta_1+17\rho_2)g^{\mu \nu}S^{\alpha}_{\alpha 2}]V_\nu^- +
	\sqrt{2}\big(\mathbf{e}_{(13)} +\nonumber
	\\
	&&i\mathbf{e}_{(14)}\big)(\beta_-S^{\mu \nu -} + \beta_-g^{ \mu \nu }S^{\alpha -}_{\alpha} + 17\rho_-g^{\mu \nu}S^{\alpha-}_{\alpha})A_\nu +\sqrt{2}\left(\mathbf{e}_{(23)} + i\mathbf{e}_{(24)}\right)(\beta_-S^{\mu \nu -} +\nonumber
	\\
	&&+ \beta_-g^{ \mu \nu }S^{\alpha -}_{\alpha}
	+ 17\rho_-g^{\mu \nu}S^{\alpha-}_{\alpha})U_\nu -4i\mathbf{e}_{[34]}(\mathbf{e}^{[+-][\mu\nu]} + \mathbf{e}^{[12][\mu\nu]})V_{\nu}^- +\sqrt{2}\big(\mathbf{e}_{[13]}+\nonumber
	\\
	&&+i\mathbf{e}_{[14]}\big)(\mathbf{e}^{[-1][\mu\nu]} + \mathbf{e}^{[-2][\mu\nu]})A_\nu+
	\sqrt{2}\left(\mathbf{e}_{[13]}+i\mathbf{e}_{[14]}\right)(\mathbf{e}^{[-1][\mu\nu]} + \mathbf{e}^{[-2][\mu\nu]})U_\nu+\nonumber
	\\
	&&+2\left( \mathbf{e}_{(33)} + \mathbf{e}_{(44)}\right)( \mathbf{e}^{(11)(\mu \nu)} + g^{\mu \nu} + \mathbf{e}^{(22)(\mu \nu)}  + \mathbf{e}^{(12)(\mu \nu)} + g^{\mu \nu}\mathbf{e}^{(12)\alpha}_\alpha + \mathbf{e}^{(+-)(\mu \nu)} +
	\\
	&&+  g^{\mu \nu} \mathbf{e}^{(+-)\alpha}_\alpha + 2g^{\mu \nu}\mathbf{e}^{(11)\alpha}_\alpha + 2g^{\mu \nu}\mathbf{e}^{(22)\alpha}_\alpha)V_\nu^- + \sqrt{2}\left( \mathbf{e}_{(13)} + i\mathbf{e}_{(14)}  \right) \big( \mathbf{e}^{(-1)(\mu \nu)} +\nonumber
	\\ 
	&&+ \mathbf{e}^{(-2)(\mu \nu)} + 17g^{\mu \nu}\mathbf{e}^{(-1)\alpha}_\alpha + 17g^{\mu \nu}\mathbf{e}^{(-2)\alpha}_\alpha \big)A_\nu+\sqrt{2}\left( \mathbf{e}_{(23)} + i\mathbf{e}_{(24)}  \right) \big( \mathbf{e}^{(-1)(\mu \nu)} +\nonumber
	\\
	&&+ \mathbf{e}^{(-2)(\mu \nu)} + 17g^{\mu \nu}\mathbf{e}^{(-1)\alpha}_\alpha + 17g^{\mu \nu}\mathbf{e}^{(-2)\alpha}_\alpha \big)U_\nu + \left( \mathbf{e}_{(33)} - \mathbf{e}_{(44)}  \right) ( \mathbf{e}^{(--)(\mu \nu)}+\nonumber
	\\
	&& + 16g^{\mu \nu}\mathbf{e}^{(--)\alpha}_\alpha )V_\nu^+
	+2i\mathbf{e}_{(34)}(\mathbf{e}^{(--34)(\mu \nu)} +16\mathbf{e}^{(--34)\alpha}_\alpha) V_\nu^+  \nonumber
\end{eqnarray}
and
\begin{eqnarray}
	&&\partial_\nu \{2a_3V^{\mu\nu+}+2\beta_+\beta_-S^{\nu\mu+}+2\big(16\rho_+\rho_-+\rho_+\beta_-+\rho_-\beta_+ \nonumber
	\\
	&&+1/4\left(\xi_{33}+\xi_{44}\right) \big)g^{\nu\mu}S_\alpha^{\alpha+}+2b_3\left(\mathbf{e}^{[+1][\mu\nu]}+\mathbf{e}^{[+2][\mu\nu]}\right)+\nonumber
	\\
	&&+\beta_-\left(\mathbf{e}^{(+1)(\nu\mu)}+g^{\nu\mu}\mathbf{e}_\alpha^{(+1)\alpha}+\mathbf{e}^{(+2)(\nu\mu)}+g^{\nu\mu}\mathbf{e}_\alpha^{(+2)\alpha}\right)+
	\\
	&&+34\rho_-g^{\nu\mu}\left(\mathbf{e}^{(+1)\alpha}_\alpha+\mathbf{e}^{(+2)\alpha}_\alpha\right) \}=J_V^{\mu+}\nonumber
\end{eqnarray}
\begin{eqnarray}
	&&J_V^{\mu+}=-2i\mathbf{e}_{[34]}\left(b_1F^{\mu\nu}+b_2U^{\mu\nu}\right)V_\nu^++\sqrt{2}\left(\mathbf{e}_{[13]}-i\mathbf{e}_{[14]}\right)V^{\mu\nu+}A_\nu \nonumber 
	\\
	&&+\big(\mathbf{e}_{(33)}+\mathbf{e}_{(44)}\big)[\beta_1S^{\mu \nu}_1 +\beta_2S^{\mu \nu}_2 +
	(\beta_1+17\rho_1)g^{\mu \nu}S^{\alpha}_{\alpha 1}
	(\beta_1+17\rho_2)g^{\mu \nu}S^{\alpha}_{\alpha 2}]V_\nu^+ +\nonumber
	\\
	&&+\sqrt{2}\left(\mathbf{e}_{(13)} - i\mathbf{e}_{(14)}\right)(\beta_+S^{\mu \nu +} + \beta_+g^{ \mu \nu }S^{\alpha +}_{\alpha} + 17\rho_+g^{\mu \nu}S^{\alpha+}_{\alpha})A_\nu+ \sqrt{2}\big(\mathbf{e}_{(23)}\nonumber
	\\
	&&- i\mathbf{e}_{(24)}\big)(\beta_+S^{\mu \nu +} + \beta_+g^{ \mu \nu }S^{\alpha +}_{\alpha} + 17\rho_+g^{\mu \nu}S^{\alpha+}_{\alpha})U_\nu -4i\mathbf{e}_{[34]}(\mathbf{e}^{[+-][\mu\nu]} + \mathbf{e}^{[12][\mu\nu]})V_{\nu}^+\nonumber
	\\
	&&+\sqrt{2}\left(\mathbf{e}_{[13]}-i\mathbf{e}_{[14]}\right)(\mathbf{e}^{[+1][\mu\nu]} + \mathbf{e}^{[+2][\mu\nu]})A_\nu+
	\sqrt{2}\left(\mathbf{e}_{[13]}-i\mathbf{e}_{[14]}\right)(\mathbf{e}^{[+1][\mu\nu]} + \mathbf{e}^{[+2][\mu\nu]})U_\nu\nonumber
	\\
	&&+2\left( \mathbf{e}_{(33)} + \mathbf{e}_{(44)}\right)( \mathbf{e}^{(11)(\mu \nu)} + g^{\mu \nu} + \mathbf{e}^{(22)(\mu \nu)} + g^{\mu \nu} + \mathbf{e}^{(12)(\mu \nu)} + g^{\mu \nu}\mathbf{e}^{(12)\alpha}_\alpha + \mathbf{e}^{(+-)(\mu \nu)} 
	\\
	&&+g^{\mu \nu} \mathbf{e}^{(+-)\alpha}_\alpha + 2g^{\mu \nu}\mathbf{e}^{(11)\alpha}_\alpha + 2g^{\mu \nu}\mathbf{e}^{(22)\alpha}_\alpha)V_\nu^+ + 
	\sqrt{2}\left( \mathbf{e}_{(13)} - i\mathbf{e}_{(14)}  \right) \big( \mathbf{e}^{(+1)(\mu \nu)} + \mathbf{e}^{(+2)(\mu \nu)} \nonumber
	\\ 
	&&+17g^{\mu \nu}\mathbf{e}^{(+1)\alpha}_\alpha + 17g^{\mu \nu}\mathbf{e}^{(+2)\alpha}_\alpha \big)A_\nu+\sqrt{2}\left( \mathbf{e}_{(23)} - i\mathbf{e}_{(24)}  \right) \big( \mathbf{e}^{(+1)(\mu \nu)}  + 17g^{\mu \nu}\mathbf{e}^{(+1)\alpha}_\alpha +\nonumber
	\\
	&&++ \mathbf{e}^{(+2)(\mu \nu)}+ 17g^{\mu \nu}\mathbf{e}^{(+2)\alpha}_\alpha \big)U_\nu+ \left( \mathbf{e}_{(33)} - \mathbf{e}_{(44)}  \right) ( \mathbf{e}^{(++)(\mu \nu)} + 16g^{\mu \nu}\mathbf{e}^{(++)\alpha}_\alpha )V_\nu^-\nonumber
	\\
	&&++2i\mathbf{e}_{(34)}(\mathbf{e}^{(++34)(\mu \nu)}+ 16\mathbf{e}^{(--34)\alpha}_\alpha) V_\nu^- \nonumber
\end{eqnarray}

A whole EM physics is established. Maxwell equations are extended to a field's strengths network. An associated and nonlinear dynamics are generated in terms of observables and free coefficients. A dualism appears. A complementarity between individual-collective fields strengths. Granular and collective fields' strengths propagate together in space-time and work as their own sources by coupling with potential fields. The coupling between fields strengths and potential fields is through adimensional free coefficients. Sources beyond electric charge.

Including the constraints equations,
\begin{equation}
	\partial_\nu S^{\nu \mu I} =\partial_\nu F^{\nu \mu I} + g^{\nu\mu}\partial_\mu S^{\alpha I}_{\alpha }
\end{equation}
and
\begin{equation}
	\partial_\nu \mathbf{e}^{(\nu\mu)} = \frac{1}{2}\mathbf{e}_{(IJ)}S^{\alpha I}_\alpha A^{\mu J }+ \mathbf{e}_{(IJ)}S^{\nu \mu I}A_{\mu}^J -\frac{1}{2}\partial_\nu g^{\nu \mu} \mathbf{e}^{\alpha}_\alpha \label{Equações subsidiarias}
\end{equation}
one suppresses the field's strengths, by mixing spin-1 and spin-0 together.

Next development is to split the above dynamics at spin-1 and spin-0 sectors. Apply the transversal and longitudinal operators, $\theta_{\mu \nu} = g_{\mu \nu} - \frac{\partial_\mu \partial_\nu}{\Box}$ and $ \omega_{\mu \nu} = \frac{\partial_\mu \partial_\nu}{\Box}$. Physically, such spin-1 and spin-0 dynamics separation is valid, due to the fact that, although non-local operators are applied, covariance is maintained. Lorentz's transformation and the electric charge symmetry preserved.

For spin-1 sector, it yields
\begin{equation}
	\partial_\nu\{ 4\left(a_1 +\beta_1 \right)F^{\nu \mu} + 2b_1\left( \mathbf{e}^{[12][\nu \mu]} + \mathbf{e}^{[+-][\nu \mu]}\right) \} = J^{\mu}_{A T},
\end{equation}
\begin{equation}
	\partial_\nu\{ 4\left(a_2 +\beta_2 \right)U^{\nu \mu} + 2b_2\left( \mathbf{e}^{[12][\nu \mu]} + \mathbf{e}^{[+-][\nu \mu]}\right) \}-2\mathbf{m}_U^2 U_L^{\mu} = J^{\mu}_{U T},
\end{equation}
\begin{equation}
	\partial_\nu\{ 2\left(a_3+\beta_+\beta_-\right)V^{\nu\mu-} + 2b_3\left( \mathbf{e}^{[-1][\nu \mu]}+\mathbf{e}^{[-2][\nu \mu]}\right) \} - \mathbf{m}_V^2V_T^{\mu-}=J^{\mu -}_{V T},
\end{equation}
\begin{equation}
	\partial_\nu\{ 2\left(a_3+\beta_+\beta_-\right)V^{\nu\mu+} + 2b_3\left( \mathbf{e}^{[+1][\nu \mu]}+\mathbf{e}^{[+2][\nu \mu]}\right) \} - \mathbf{m}_V^2V_T^{\mu +}=J^{\mu +}_{V T}. \label{V spin-1}
\end{equation}
where
\begin{equation}
	J_{I T}^{ \mu} = \theta^{\mu \nu}J_{\nu I}
\end{equation}

For spin-0 sector:
\begin{eqnarray}
	&&\partial_\nu \{ 4\left(\beta_1 + \rho_1\beta_1+11\rho_1 + \frac{1}{2}\xi_{(11)} \right)S_{\alpha}^{\alpha 1} + 2\big( \rho_1 \beta_2 + \nonumber
	\\
	&&\rho_2 \beta_1 + 22\rho_1\rho_2 +1/2\xi_{(12)}\big)S_\alpha^{\alpha 2} + \big( 34\rho_1+ 
	\\
	&&- 2\beta_1\big)g^{\nu \mu}\left( \mathbf{e}^{(11)\alpha}_\alpha + \mathbf{e}^{(22)\alpha}_\alpha + \mathbf{e}^{(12)\alpha}_\alpha + \mathbf{e}^{(+-)\alpha}_\alpha \right) \} = J^{\mu}_{A L},\nonumber
\end{eqnarray}
\begin{eqnarray}
	&&\partial_\nu \{ 4\left(\beta_2 + \rho_2\beta_1+11\rho_2 +\frac{1}{2}\xi_{(22)} \right)S_{\alpha}^{\alpha 2} + 2\big( \rho_1 \beta_2 + \rho_2 \beta_1 + \nonumber
	\\
	&&22\rho_1\rho_2 +\frac{1}{2}\xi_{(12)}\big)S_\alpha^{\alpha 1} 
	+ \left( 34\rho_2 - 2\beta_2\right)g^{\nu \mu}\big( \mathbf{e}^{(11)\alpha}_\alpha \nonumber
	\\
	&&+ \mathbf{e}^{(22)\alpha}_\alpha + \mathbf{e}^{(12)\alpha}_\alpha + \mathbf{e}^{(+-)\alpha}_\alpha \big) \}-2\mathbf{m}_U^2 U_L^{\mu} = J^{\mu}_{U L},\nonumber
\end{eqnarray}
\begin{multline}
	\partial_{\nu} \{ 2\left( \beta_+\beta_- + 16\rho_+\rho_- + \rho_+\beta_- + \rho_-\beta_+ +\frac{1}{4}\left(\xi_{(33)} + \xi_{(44)} \right)\right)g^{\nu \mu}S_\alpha^{\alpha -}\\
	+ \left( 34\rho_+ - 2\beta_+\right)g^{\nu \mu}\left(\mathbf{e}^{(-1) \alpha}_\alpha + \mathbf{e}^{(-2)\alpha}_\alpha\right)   \} - \mathbf{m}_V^2 V^{\mu -}_L = J^{\mu -}_{V L},
\end{multline}
\begin{multline}
	\partial_{\nu} \{ 2\left( \beta_+\beta_- + 16\rho_+\rho_- + \rho_+\beta_- + \rho_-\beta_+ \frac{1}{4}\left(\xi_{(33)} + \xi_{(44)} \right)\right)g^{\nu \mu}S_\alpha^{\alpha +}\\
	+ \left( 34\rho_- - 2\beta_-\right)g^{\nu \mu}\left(\mathbf{e}^{(+1) \alpha}_\alpha + \mathbf{e}^{(+2)\alpha}_\alpha\right)   \} - \mathbf{m}_V^2 V^{\mu +}_L = J^{\mu +}_{V L}.
\end{multline}
where
\begin{equation}
	J_{I L}^{\mu} = \omega^{\mu \nu}J_{\nu I}
\end{equation}

Thus, although the Lorentz group representation $A_{\mu I} \in (\frac{1}{2},\frac{1}{2})$ contains spin-0 and spin-1, dynamically, they are distinct. A symmetry pluriform is expressed. Splitting at LHS equations, it yields vector and a scalar kind of photons with different quanta. However, at RHS, the mixing spin terms are preserved. Notice that the granular spin-0 propagation can be controlled by gauge fixing parameters.

section{Electric charge symmetry}

A generic electric charge physics based on two phenomenologies is being studied. It englobes the mutation and conservation properties. Macroscopically, Maxwell understood the electric charge conservation. A microscopic interpretation is expected based on four intermediate fields. Consider the electric charge set $\{+,0,-\}$ transformations under the quadruplet $\{A_\mu, U_\mu, V_\mu^{\pm}\}$. Connect these fields under a modified electric charge symmetry.

The electric charge nature is extended. At $18^{th}$ century, Benjamin Franklin identified positive and negative electric charges and formulated the electric charge conservation. The microscopic charge transfer phenomena introduce a new feature for electric charge. It is the physical presence of a neutral charge. Considering that electric charge is a generator of $U_{Q}(1)$ global gauge group, the corresponding quantum number is continuous. Therefore, the electric charge value, $q=N_{q}e$, can take any value as an integer, fractional, or even zero. A formulation enclosing a neutral electric charge.

There is physics beyond Maxwell's electric charge properties. The corresponding continuity equation and coupling constant be enlarged. The gauge parameter is related to eqs. (\ref{Usual photon}-\ref{Charge photon}) develops the quadruplet association $ U_{Q} \equiv U(1) \times SO(2)_{global}$. This new symmetry produces a whole EM, expressed through Lagrangian and Noether theorem. Three identities are obtained from Lagrangian invariance,
\begin{equation}
	\alpha\partial_\mu J_N^{\mu} + \partial_{\nu}\alpha\{\partial_\mu K^{\mu \nu} + J_N^{\nu}\} + \partial_\mu \partial_{\nu} \alpha K^{\mu \nu} = 0 
\end{equation}

The first one corresponds to the electric charge conservation,
\begin{equation}
	\partial_\mu J^{\mu}_N =0 \label{eq. continuity}
\end{equation}
with
\begin{equation}
	J^\mu_N = iq  \left\{ \left[V^+_\nu \frac{\partial L}{\partial\left(\partial_\nu V_\mu^{+}\right)} \right]  - \left[V^-_\nu \frac{\partial L}{\partial\left(\partial_\nu V_\mu^{-}\right)} \right] \right\} 
\end{equation}
It gives,
\begin{multline}\label{Corrente Noether}
	J_N^\mu = -iq\{  4a_3\left(  V_\nu^+V^{\nu \mu -} -V_\nu^-V^{\nu \mu +} \right) + 4\beta_+\beta_- \left( V_\nu^+S^{\nu \mu -} - V_\nu^-S^{\nu \mu+} \right) +\\
	+ 4\left(16\rho_+\rho_- + \rho_+\beta_-+\rho_-\beta_+ \right)g^{\nu \mu}\left( V^+_\nu S_\alpha^{\alpha-} - V^{-}_\nu S_\alpha^{\alpha +} \right)  +4[b_3 V_\nu^+\left(\mathbf{e}^{[-1][\nu \mu]} + \mathbf{e}^{[-2][\nu \mu]}\right) + \\
	-b_3V_\nu^-\left(\mathbf{e}^{[+1][\nu \mu]} + \mathbf{e}^{[+2][\nu \mu]}\right) ] + 4\beta_+V_\nu^+\left( \mathbf{e}^{(-1)(\nu\mu)}+g^{\nu\mu}\mathbf{e}^{(-1)\alpha}_\alpha+\mathbf{e}^{(-2)(\nu\mu)}+g^{\nu\mu}\mathbf{e}^{(-2)\alpha}_\alpha \right)+\\
	-4\beta_-V_\nu^-\left( \mathbf{e}^{(+1)(\nu\mu)}+g^{\nu\mu}\mathbf{e}^{(+1)\alpha}_\alpha+\mathbf{e}^{(+2)(\nu\mu)}+g^{\nu\mu}\mathbf{e}^{(+2)\alpha}_\alpha \right)
	+17\rho_+V_\nu^+g^{\nu\mu}\left(\mathbf{e}^{(-1)\alpha}_\alpha+ \mathbf{e}^{(-2)\alpha}_\alpha \right)+\\
	-17\rho_-V_\nu^-g^{\nu\mu}\left(\mathbf{e}^{(+1)\alpha}_\alpha+ \mathbf{e}^{(+2)\alpha}_\alpha \right)
	\} 
\end{multline} 
Eq. (\ref{Corrente Noether}) reproduces the electric charge coupling between positive and negative charges. The first novelty is electric charge coupling with observable beyond Maxwell. It is also not restricted to charged fields. Chargeless fields are included through collective fields, as term $\mathbf{e}^{[+1][\mu \nu]}$ shows. 

The second Noether identity is the symmetry equation. It is identified as the electric charge equation,
\begin{equation}
	\partial_{\nu} K^{\nu \mu} + J^{\mu}_N =0\label{Noether charge}
\end{equation}
where
\begin{equation} \label{k equation}
	K^{\nu \mu} = k_1\frac{\partial L}{\partial\left(\partial_\nu A_\mu \right)} + k_2\frac{\partial L}{\partial\left(\partial_\nu U_\mu\right)} + k_+\frac{\partial L}{\partial\left(\partial_\nu V_\mu^{+}\right)} + k_-\frac{\partial L}{\partial\left(\partial_\nu V_\mu^{-}\right)} 
\end{equation}
which gives the following gauge-invariant equation,
\begin{eqnarray}\label{equação da simetria}
	&&\partial_\nu \{4k_1a_1F^{\nu \mu} + 4k_2a_2U^{\nu \mu} + 2k_-a_3V^{\nu \mu +} +2k_+a_3V^{\nu \mu -} +\nonumber
	\\
	&&+4(11k_1\rho_1 + \frac{1}{2}k_2\rho_2\beta_1+ 11k_2\rho_1\rho_2 +\frac{1}{4}k_2\xi_{(12)})g^{\nu \mu}S_{\alpha}^{\alpha 1 }+\nonumber
	\\
	&&+4(11k_2\rho_2 + \frac{1}{2}k_1\rho_2\beta_1 +
	11k_1\rho_1\rho_2 +\frac{1}{4}k_1\xi_{(12)})g^{\nu \mu}S^{\alpha}_{\alpha 2}
	\\
	&&+2\left(b_1 k_1 + b_2 k_2\right)\mathbf{e}^{[+-][\nu \mu]} \} = -J_N^{\mu}.\nonumber 
\end{eqnarray}

Eq.(\ref{equação da simetria}) establishes the electric charge flow through a quadruplet. It shows electric charge as a source for propagating diverse granular and collective fields strengths. The new behavior is produced on its flux features given by conduction, transmission, conservation, interaction. Conducted by fields quadruplet, transmitted through positive, negative, zero charges, conserved by gauge symmetry, and interacting with potential fields coupling with fields strengths.

Appendix B introduces a complementary consideration to eq. (\ref{k equation}). Two distinct electric charge equations are derived by applying the transversal and longitudinal operators. Electric charge equation is separated into spin-1 and spin-0 sectors.

For spin-1:
\begin{equation}
	\partial_\nu \{4k_1a_1F^{\nu \mu} + 4k_2a_2U^{\nu \mu} + 2k_+a_3V^{\nu \mu +} +2k_-a_3V^{\nu \mu +} +2\left(b_1 k_1 + b_2 k_2\right)\mathbf{e}^{[+-][\nu \mu]} =-J_{N T}^{\mu}
\end{equation}

For spin-0:
\begin{multline}
	\partial_{\nu} \{  2\left( 22k_1\rho_1 + k_2\rho_2\beta_1 + \frac{1}{2}k_2\xi_{(12)}+ 22k_2\rho_1\rho_2 \right)g^{\mu \nu} S^\alpha_{\alpha 1} + \\
	+ 2\left(k_1\rho_1\beta_2 + 22k_1\rho_1\rho_2 + \frac{1}{2}\xi_{(12)} + 22k_2\rho_2\right)g^{\nu \mu}S^\alpha_{\alpha 2} \} = -J_{N L}^{\mu}
\end{multline}
where $J^{\mu}_{N T} = \theta^\mu_\nu J^{\nu}_N$ and $J^{\mu}_{N L} = \omega^{\mu}_\nu J^{\nu}_N$

The third Noether identity is the constraint
\begin{equation}
	\partial_\mu\partial_{\nu}K^{(\nu \mu)} = 0
\end{equation}
Considering that a symmetric tensor $K_{(\mu \nu)}$ can be separated as
\begin{equation}
	K_{(\mu \nu)} = \tilde{K}_{(\mu \nu)} +g_{\mu \nu}K_\alpha^\alpha
\end{equation}
where $\tilde{K}_{(\mu \nu)}$ is a tensor with trace zero, it gives 

\begin{equation}
	\Box K_\alpha^\alpha + \partial^\mu\partial^\nu \tilde{K}_{(\mu \nu)} =0 \label{K identidade}
\end{equation}
Eq. (\ref{K identidade}) shows that Noether constraint does not interfere with Lagrangian physicality. No term at eq. (\ref{Campo granular primeiro}) is suppressed by the constraint. The presence of granular and collective fields strengths is preserved. 

The quadruplet electric charge symmetry enlarges the usual electric charge electromagnetic scenario. It replaces Maxwell to an $U(1) \times SO(2)_{global}$ symmetry from where the potential fields are tied by electric charge. The experimental conserved charge [54] relationship is reproduced under four fields associations, as eq. (\ref{eq. continuity}) shows. However, interactions not necessarily related to electric charge coupling appear. The quadruplet mutation physics introduces that the electric charge is no more a universal coupling constant. Consequently, the beta decay reaction may be reinterpreted as a phenomenon belonging to electric charge transfer. The Fermi coupling constant dimension is indicating the presence of a massive propagator, and so, instead, is interpreted as weak interaction to consider as an extended EM set intermediate by massive bosons.

\section{Electromagnetic whole dynamics}
Associative physic is being expressed. A quadruplet interconnected dynamics is formed. Whole equations are constituted. A set of equations is derived combining minimal action principle, identities equations, and electric charge equation. A nonlinear EM whole dynamic is generated.

Separating the spin-1 and spin-0 sectors, one gets
For $A_{\mu T}$,
\begin{multline} 
	\partial_\nu \{4\left(a_1 + \beta_1 + k_1a_1\right)F^{\nu \mu} + 2b_1 [\mathbf{e}^{[12][\nu \mu]} + (1+ k_1)\mathbf{e}^{[+-][\nu \mu]}] \} = M^{A \mu}_{I T} + J^\mu_{A T} - J^\mu_{N T} - k_2J^\mu_{U T}  \label{Effective A}
\end{multline}
where the interactive mass term is
\begin{equation}
	M_{I T}^{A \mu} =-2 k_2\mathbf{m}_U^2 U^{\mu} -k_-\mathbf{m}_V^2V^{\mu + } -k_+\mathbf{m}_V^2V^{\mu -} \label{Massa interativa A}
\end{equation}
Eq.(\ref{Effective A}) extends the usual massless photon equation.  Eq.(\ref{Massa interativa A}) shows that although the photon is massless, it can have a penetration length. $M_{I T}^{A \mu}$ works as a source to superconductivity [55].

For $U_{\mu T}$,
\begin{eqnarray}
	\partial_\nu \{4\left(a_2 + \beta_2 + k_2a_2\right)U^{\nu \mu} + 2b_2 [\mathbf{e}^{[12][\nu \mu]} + (1+k_2)\mathbf{e}^{[+-][\nu \mu]}]  \} -2\mathbf{m}_U^2U_T^\mu = M^{A \mu}_{I T} + J^\mu_{U T}\\
	- J^\mu_{N T} - k_1J^\mu_{A T} \label{Effective U}
\end{eqnarray}
with the interactive mass term
\begin{equation}
	M^{U \mu}_I = -k_-\mathbf{m}_V^2 V^{\mu +} - k_+\mathbf{m}_V^2 V^{\mu -} \label{Massa Interativa U}
\end{equation}

Eq. (\ref{Effective U}) introduces a photon with mass [56] and eq. (\ref{Massa Interativa U}) its correspondent mass source

For $V^{\pm}_{\mu T}$,
\begin{equation}
	\partial_\nu\{ 2\left(a_3+\beta_+\beta_-\right)V^{\nu\mu-} + 2b_3\left( \mathbf{e}^{[-1][\nu \mu]}+\mathbf{e}^{[-2][\nu \mu]}\right) \} - \mathbf{m}_V^2V_T^{\mu-}=J^{\mu -}_{V T} \label{Effective V+}
\end{equation}
and
\begin{equation}
	\partial_\nu\{ 2\left(a_3+\beta_+\beta_-\right)V^{\nu\mu+} + 2 b_3 \left( \mathbf{e}^{[+1][\nu \mu]}+\mathbf{e}^{[+2][\nu \mu]}\right) \} - \mathbf{m}_V^2V_T^{\mu +}=J^{\mu +}_{V T} \label{Effective V-}
\end{equation}
Eqs. (\ref{Effective V+}-\ref{Effective V-}) are including charged vector bosons [13].

The longitudinal equations are

For $A_{\mu L}$,
\begin{equation}
	\partial_\nu \{s_{11}g^{\nu \mu}S^{\alpha}_{\alpha 1}+ c_{11}g^{\nu \mu}\left( \mathbf{e}^{(11)\alpha}_\alpha + \mathbf{e}^{(22)\alpha}_\alpha + \mathbf{e}^{(12)\alpha}_\alpha + \mathbf{e}^{(+-)\alpha}_\alpha\right)\} = J^{\mu}_{AL} - J^{\mu}_{N L} + t_{11}\mathbf{m}^2_U U^\mu_L  + \frac{1}{2}t_{11}J^\mu_{U L} \label{Equação Longitudinal A}
\end{equation}

For $U_{\mu L}$,
\begin{equation}
	\partial_\nu \{s_{22}g^{\nu \mu}S^{\alpha}_{\alpha 2}+ c_{22}g^{\nu \mu}\left( \mathbf{e}^{(11)\alpha}_\alpha + \mathbf{e}^{(22)\alpha}_\alpha + \mathbf{e}^{(12)\alpha}_\alpha + \mathbf{e}^{(+-)\alpha}_\alpha\right)\}- 2\mathbf{m}^2_U U^\mu_L  = J^{\mu}_{AL} - J^{\mu}_{N L} + t_{22}J^\mu_{U L}
\end{equation}

For $V^{\pm}_{\mu L}$,
\begin{multline}
	\partial_{\nu} \{ 2\left( \beta_+\beta_- + 16\rho_+\rho_- + \rho_+\beta_- + \rho_-\beta_+\right)g^{\nu \mu}S_\alpha^{\alpha -} + \frac{1}{2}\left(\xi_{(33)} + \xi_{(44)} \right)g^{\nu \mu}S_{\alpha}^{\alpha -} +\\
	+ \left( 34\rho_+ - 2\beta_+\right)g^{\nu \mu}\left(\mathbf{e}^{(-1) \alpha}_\alpha + \mathbf{e}^{(-2)\alpha}_\alpha\right)   \} - \mathbf{m}_V^2 V^{\mu -}_L = J^{\mu -}_{V L},
\end{multline}
and
\begin{multline}
	\partial_{\nu} \{ 2\left( \beta_+\beta_- + 16\rho_+\rho_- + \rho_+\beta_- + \rho_-\beta_+ + \frac{1}{2}\xi_{(33)} + \frac{1}{2}\xi_{(44)}\right)g^{\nu \mu}S_\alpha^{\alpha +}\\
	+ \left( 34\rho_- - 2\beta_-\right)g^{\nu \mu}\left(\mathbf{e}^{(+1) \alpha}_\alpha + \mathbf{e}^{(+2)\alpha}_\alpha\right)   \} - \mathbf{m}_V^2 V^{\mu +}_L = J^{\mu +}_{V L}.
\end{multline} 

Three kinds of EM observables appear. They are potential fields conglomerates, granular and collective fields strengths. A whole dynamic is developed. Fields strengths are derived under individual-collective dynamical correspondence at space-time and act as their own sources by coupling with potential fields. The spin-1 and spin-0 sectors embedded at Minkowski fields $A_{\mu I} \equiv (\phi_I, \vec{A}_I)$ are separated by non-local operators preserving covariance.
A diverse dynamic for spin-1 and spin-0 sectors is proposed. Similarly, in supersymmetry, one gets correlated spin-1 and spin-0 particles with different masses. They are physically viable but with a dynamic distinct for each spin- sector. The above equations are also expressing the four distinct scalar photons inserted at the Lorentz group.

A systemic relativistic equation is expressed. It provides a new EM energy manifestation. New aspects on a mass, electric charge, and couplings are derived. Besides mass without Higgs and couplings between electric and magnetic fields without electric charge, the quadruplet equations of motion express three kinds of fields associations. They are collective fields, London, and conglomerates terms. The last two are written respectively as $L = A_\mu A^{\mu}$ and $M_{IJ} = A_{\mu I}A^{\mu}_J$. They introduce the following terms at equations
\begin{equation}
	l_I^\mu \equiv gLA_I^\mu, \ L \equiv A_{\nu J}A^\nu_J \label{London terms}
\end{equation}
and
\begin{equation}
	c_I^\mu \equiv g_{J K}M_{J K}A_I^\mu, \ M_{J K} \equiv A_{\nu J}A^\nu_K, \label{conglomerates terms}
\end{equation}

The current term is
\begin{equation}
	j_I^\mu \equiv f_I X^{\mu \nu}_JA_{\nu J} \label{correntes efetivas}
\end{equation}
where $A_{\mu I}$ is any field belonging to the quadruplet $A_{\mu I} = \{A_\mu, U_\mu, V^{\pm}_\mu\}$ and $X^{\mu \nu}_I$ any granular or collective field strength. For the particular case where eq.(\ref{correntes efetivas}) is expressed as
\begin{equation}
	j^{\mu s}_I= f_J \partial_{\nu} \left(A_I^{\nu}A^{\mu}_J - A_J^{\nu}A^{\mu}_I\right)
\end{equation}
it can be rewritten in terms of the spin operator.
\begin{equation}
	j^{\mu s}_I = f_I i \partial_{\nu}\left( A_J^{\alpha}\left(\Sigma^{\nu \mu}\right)_{\alpha}A^{\beta}_J\right) \label{Spin terms}
\end{equation}
where $\left(\Sigma_{\alpha \beta}\right)^{\mu}_{\nu} = -i \left(\delta^{\mu}_{\alpha}\eta_{\alpha \beta} - \eta_{\alpha \nu}\delta^{\mu}_{\rho} \right)$.

Maxwell's limitations are surrounded. The quadruplet dynamics crosses consistently the Maxwell frontier. A nonlinear EM appears. Potential fields are expressed explicitly in the equations. Granular and collective fields strengths are developed. The collective fields' strengths are identified to polarization and magnetization vectors [57], Eq. (\ref{London terms}) to London superconductivity term [58], eq. (\ref{conglomerates terms}) to Anderson plasmon [59]. Photon-photon interaction is revived [60]. Eq. (\ref{Spin terms}) incorporates spin presence [61]. 

Electric charge properties are derived. Historically, physics development has been showing diverse performances on electric charge. The renormalization group equation introduced the value of structure constant depending on momentum, the standard model modulates its value, quarks introduce a fractional value, Gell-Mann-Nishijima on quantum numbers [62]. A further feature is electric charge mutation. Writing a generic current, one gets
\begin{eqnarray}
	&&J^{\mu} \sim 2 b_1 \mathbf{e}_{[12]}F^{\mu \nu}A_{\nu} + 4\mathbf{e}_{[12]}\mathbf{e}^{[+-][\mu \nu]} A_\nu + \left(\mathbf{e}_{[13]} - i \mathbf{e}_{[14]} + qa_3\right)V^{\mu \nu+}V_{\nu -} \nonumber
	\\
	&&+ \left(\mathbf{e}_{[13]} - i \mathbf{e}_{[14]} + q\right)\mathbf{e}^{[+1][\mu \nu]}V_\nu^+  \label{Corrente expressa}
\end{eqnarray}
where eq.(\ref{Corrente expressa}) shows diverse types of EM coupling beyond electric charge. Showing that EM interactions are not restricted to electric charge. Physically, the EM flux comes first, and it includes neutral couplings.

Thus, an extended EM is proposed. Given the above features, the whole quadruplet dynamics are derived. It gives,

For photon field:

$A_{\mu}$ spin-1:
\begin{eqnarray}
	&&\partial_\nu \{4\left(a_1 + \beta_1 + k_1a_1\right)F^{\nu \mu} + 2b_1[\mathbf{e}^{[12][\nu \mu]} + (1+k_1)\mathbf{e}^{[+-][\nu \mu]}] \} +\nonumber
	\\
	&& + l_{A T}^\mu + c_{A T}^{\mu} =   M^{A\mu}_{IT}+ j^\mu_{A T}- J^\mu_{N T}  - k_2J^{\mu}_{U T}
\end{eqnarray}

$A_{\mu}$ spin-0:
\begin{eqnarray}
	&&\partial_\nu \{s_{11}g^{\nu \mu}S^{\alpha}_{\alpha 1}+ c_{11}g^{\nu \mu}\left( \mathbf{e}^{(11)\alpha}_\alpha + \mathbf{e}^{(22)\alpha}_\alpha + \mathbf{e}^{(12)\alpha}_\alpha + \mathbf{e}^{(+-)\alpha}_\alpha\right)\} \nonumber
	\\
	&&+ l_{A L}^\mu + c_{A L}^{\mu}  = j^{\mu}_{A L} - J^{\mu}_{N L} + t_{11}\mathbf{m}_U U^{\mu}_{L} +\frac{1}{2}t_{11}J^\mu_{U L}
\end{eqnarray}
where
\begin{eqnarray}
	&&l_A^\mu = - A_\nu \big\{ 4\mathbf{e}_{(11)}\big( \mathbf{e}^{(+-)(\mu \nu)} + 2g^{\mu \nu}\mathbf{e}^{(11)\alpha}_{\alpha} + g^{\mu \nu}\mathbf{e}^{(22)\alpha}_\alpha + 2g^{\mu \nu}\mathbf{e}^{(11)\alpha}_{\alpha} + 17g^{\mu\nu} \mathbf{e}^{(11)\alpha}_{\alpha}\big) \nonumber
	\\
	&&+ 4\mathbf{e}_{(12)}\mathbf{e}^{(11)(\mu \nu)}\big\} 
	- U_\nu \{4 \mathbf{e}_{[12]}\mathbf{e}^{[12][\mu \nu]}  +4\mathbf{e}_{(12)}( 17 g^{\mu \nu}\mathbf{e}^{(11)\alpha}_{\alpha} + 2\mathbf{e}^{(22)(\mu \nu)}  +g^{\mu \nu}\mathbf{e}^{(12)\alpha}_\alpha 
	\\
	&&+ 18g^{\mu \nu}\mathbf{e}^{(22)\alpha}_\alpha)\},\nonumber
\end{eqnarray}
\begin{eqnarray}
	&&c^\mu_A = -A_\nu\{ 4\mathbf{e}_{(11)}(\mathbf{e}^{(+-)(\mu \nu)} + 17g^{\mu \nu}\mathbf{e}^{(+-)\alpha}_{\alpha}) \} - U_\nu\{ 4\mathbf{e}_{[12]}\mathbf{e}^{[+-][\mu \nu]} + 72\mathbf{e}_{(12)}g^{\mu \nu}\mathbf{e}^{(+-)\alpha}_{\alpha}\}+\nonumber
	\\
	&&-V_\nu^-\{ \sqrt{2}\left(\mathbf{e}_{[13]} - i\mathbf{e}_{[14]}\right)(\mathbf{e}^{[+1][\mu \nu]} + \mathbf{e}^{[+2][\mu \nu]}) + \sqrt{2}\left(\mathbf{e}_{(13)} - i\mathbf{e}_{(14)}\right)(\mathbf{e}^{(+1)(\mu \nu)} 
	+  \nonumber
	\\
	&&+18g^{\mu \nu}\mathbf{e}^{(+1)\alpha}_{\alpha}+\mathbf{e}^{(+2)(\mu \nu)} + 18g^{\mu \nu}\mathbf{e}^{(+2)\alpha}_{\alpha})\}
	-V_\nu^+\{ \sqrt{2}\left(\mathbf{e}_{[13]} + i\mathbf{e}_{[14]}\right)(\mathbf{e}^{[-1][\mu \nu]}   +
	\\
	&&+\mathbf{e}^{[-2][\mu \nu]})+\sqrt{2}\left(\mathbf{e}_{(13)} + i\mathbf{e}_{(14)}\right)(\mathbf{e}^{(-1)(\mu \nu)}+ 18g^{\mu \nu}\mathbf{e}^{(-1)\alpha}_{\alpha} +\mathbf{e}^{(-2)(\mu \nu)} + 18g^{\mu \nu}\mathbf{e}^{(-2)\alpha}_{\alpha}\}\nonumber
\end{eqnarray}
The current term is
\begin{eqnarray}
	&&j^{\mu}_A = A_\nu\{ 2\mathbf{e}_{(11)}\big[( \frac{1}{2}\beta_1 + 17\rho_1) g^{\mu \nu}S^{\alpha}_{\alpha 1} +  \left( \beta_2 + 17\rho_2 \right) g^{\mu \nu}S^{\alpha}_{\alpha 2} + \beta_1S^{\mu \nu}_1\big]  -\beta_1\mathbf{e}_{(12)}g^{\mu \nu}S^{\alpha}_{\alpha 2}\nonumber
	\\
	&&+ \beta_1\mathbf{e}_{(22)}S^{(\mu \nu)}_{2} \} + U_\nu\{  2\mathbf{e}_{[12]}\left(b_1 F^{\mu \nu}+ b_2 U^{\mu \nu}\right) + 2\mathbf{e}_{(12)}\big[ \beta_2 S^{\mu \nu}_2 + \left(\beta_1 +17\rho_1 \right)g^{\mu \nu}S^{\alpha}_{\alpha 1}\nonumber
	\\
	&& +  \left( \beta_2 + 17\rho_2 \right) g^{\mu \nu}S^{\alpha}_{\alpha 2}  \big]  -\beta_1\mathbf{e}_{(22)}g^{\mu \nu}S^{\alpha}_{\alpha 2} - \beta_1\mathbf{e}_{(22)}S^{\mu \nu}_2 \}  
	+V_\nu^+\big\{ \sqrt{2}\left( \mathbf{e}_{[13]} +i \mathbf{e}_{[14]} \right)V^{\mu \nu -}\nonumber 
	\\
	&&+ \sqrt{2} \left( \mathbf{e}_{(13)} +i \mathbf{e}_{(14)} \right)\left( \beta_-S^{\mu \nu -} + 17\rho_-g^{\mu \nu} S^{\alpha -}_{\alpha } +  \right) -\beta_1\left(\mathbf{e}_{(33)} + \mathbf{e}_{(44)} \right) \big(g^{\mu \nu}S_\alpha^{\alpha -}   +
	\\
	&&+2 S^{\mu \nu -}  \big)\big\} +V_\nu^-\{ \sqrt{2}\left( \mathbf{e}_{[13]} -i \mathbf{e}_{[14]} \right)V^{\mu \nu +} -\beta_1\left(\mathbf{e}_{(33)} + \mathbf{e}_{(44)} \right) \left(g^{\mu \nu}S_\alpha^{\alpha +}  +2 S^{\mu \nu +}  \right)\nonumber
	\\
	&&+ \sqrt{2} \left( \mathbf{e}_{(13)} -i \mathbf{e}_{(14)} \right)\left( \beta_+S^{\mu \nu +} + 17\rho_+g^{\mu \nu} S^{\alpha +}_{\alpha } + \beta_+g^{\mu \nu} S^{\alpha +}_{\alpha }  \right)\} \nonumber
\end{eqnarray}

For massive photon field:

$U_{\mu}$ spin-1:
\begin{eqnarray}
	&&\partial_\nu \left\{ 4\left(a_2 +\beta_2 + a_2k_2\right)U^{\nu \mu}  + 2b_2 [\mathbf{e}^{[12][\nu \mu]} + (1+k_2)\mathbf{e}^{[+-][\nu \mu]}] + 2b_2k_2\mathbf{e}^{[+-][\nu \mu]}\right\}+\nonumber
	\\
	&& -2\mathbf{m}_U^2 U^\mu_T + l^{\mu}_{U T} + c^\mu_{U} = M^{U \mu}_{I T} + j^{\mu}_{U T} - k_1 J^{\mu}_{A T} - J^\mu_{N T}
\end{eqnarray}

$U_{\mu}$ spin-0:
\begin{multline}
	\partial_{\nu} \left\{ 4s_{22}g^{\mu \nu}S^{\alpha}_{\alpha 2} + c_{22}g^{\mu \nu}\left(\mathbf{e}^{(11)\alpha}_\alpha + \mathbf{e}^{(22)\alpha}_\alpha + \mathbf{e}^{(12)\alpha}_\alpha + \mathbf{e}^{(11)\alpha}_\alpha\right)\right\} - 2\mathbf{m}_U^{2} U^{\mu}_L +l^{\mu}_{U L} +c^{\mu}_{U L}  = \\
	=  j^{\mu}_{U L} - k_1 J^{\mu}_{A L} -J^{\mu}_{N L}
\end{multline}
where
\begin{eqnarray}
	&&l^\mu_{U} = -A_\nu \{ 4\mathbf{e}_{[12]}\mathbf{e}^{[12][\mu \nu]} + 4\mathbf{e}_{(12)}\mathbf{e}^{(22)(\mu \nu)} + 8\mathbf{e}_{(12)}\mathbf{e}^{(12)(\mu \nu)} \nonumber
	\\
	&&+ 72\mathbf{e}_{(12)}g^{\mu \nu}\mathbf{e}^{(11)\alpha}_\alpha + 4\mathbf{e}_{(12)}g^{\mu \nu}\mathbf{e}^{(12)\alpha}_\alpha + 68\mathbf{e}_{(12)}g^{\mu \nu}\mathbf{e}^{(22)\alpha}_\alpha\}\nonumber 
	\\
	&&- U_\nu\{4\mathbf{e}_{(22)}\mathbf{e}^{(22)(\mu \nu)} + 8\mathbf{e}_{(22)}g^{\mu \nu}\mathbf{e}^{(22)\alpha}_\alpha + 4\mathbf{e}_{(22)}\mathbf{e}^{(11)(\mu \nu)} 
	\\
	&&+\mathbf{e}_{(22)}g^{\mu \nu}\mathbf{e}^{(11)\alpha}_\alpha+4\mathbf{e}_{(12)}\mathbf{e}^{(12)(\mu \nu)} + 72\mathbf{e}_{(22)}g^{\mu \nu}\mathbf{e}^{(12)\alpha}_\alpha \},\nonumber
\end{eqnarray}
\begin{eqnarray}
	&&c^\mu_U = -A_\nu\{ 4\mathbf{e}_{[12]}\mathbf{e}^{[+-][\mu \nu]} + 72\mathbf{e}_{(12)}g^{\mu \nu}\mathbf{e}^{(+-)\alpha}_\alpha\} - U_\nu\{4\mathbf{e}_{(22)}\mathbf{e}^{(+-)\alpha}_\alpha\nonumber
	\\
	&&+4\mathbf{e}_{(22)}\mathbf{e}^{(+-)(\mu \nu)} + 68\mathbf{e}_{(22)}g^{\mu \nu}\mathbf{e}^{(+-)\alpha}_\alpha\}+ V_\nu^{-}\{\sqrt{2}\left(\mathbf{e}_{[23]}-i\mathbf{e}_{[24]}\right)\big(\mathbf{e}^{[+2][\mu \nu]} \mathbf{e}^{[+1][\mu \nu]} \big) \nonumber
	\\
	&&\sqrt{2}\left(\mathbf{e}_{(23)} - i\mathbf{e}_{(24)}\right)\big(\mathbf{e}^{(+1)(\mu \nu)}+ \mathbf{e}^{(+2)(\mu \nu)}+18g^{\mu \nu}\mathbf{e}^{(+1)\alpha}_\alpha +18g^{\mu \nu}\mathbf{e}^{(+2)\alpha}_\alpha\big) \}\nonumber
	\\
	&& V_\nu^+\{\sqrt{2}\left(\mathbf{e}_{[23]}+i\mathbf{e}_{[24]}\right)\mathbf{e}^{[-2][\mu \nu]} + \sqrt{2}\left(\mathbf{e}_{[23]}+i\mathbf{e}_{[24]}\right)\mathbf{e}^{[-1][\mu \nu]} 
	\\
	&&+\sqrt{2}\left(\mathbf{e}_{(23)} + i\mathbf{e}_{(24)}\right)\mathbf{e}^{(-1)(\mu \mu)}+18\sqrt{2}\left(\mathbf{e}_{(23)}+i\mathbf{e}_{(24)}\right)g^{\mu \nu}\mathbf{e}^{(-1)\alpha}_\alpha \nonumber 
	\\
	&&+\sqrt{2}\left(\mathbf{e}_{(23)}+i\mathbf{e}_{(24)}\right)\mathbf{e}^{(-2)(\mu \nu)}+ 18\sqrt{2}\left(\mathbf{e}_{(23)}+i\mathbf{e}_{(24)}\right)\mathbf{e}^{(-2)\alpha}_\alpha\}\nonumber
\end{eqnarray}
The current term is
\begin{eqnarray}
	&&j^{\mu}_U = A_\nu\{\mathbf{e}_{[12]}\left(b_1 F^{\mu \nu} + b_2U^{\mu \nu}\right) + 2\mathbf{e}_{(12)}[ (\beta_1 + 17\rho_1)g^{\mu \nu}S^{\alpha}_{\alpha 1}\nonumber
	\\
	&&+ \left( 1/2\beta_2 + 17\rho_2\right)g^{\mu \nu}S^{\alpha}_{\alpha}+\beta_1S^{\mu \nu}_1\big]-\beta_2\mathbf{e}_{(12)}g^{\mu \nu}S^{\alpha}_{\alpha 1} - 2\beta_2\mathbf{e}_{(11)}S^{\mu \nu}_1 \} + \nonumber
	\\
	&&+U_\nu\{ 2\mathbf{e}_{(22)}[\beta_1S^{\mu \nu}_1 + (\beta_1 + 17\rho_1)g^{\mu \nu}S^{\alpha}_{\alpha 1} +\left( 1/2\beta_2 + 17\rho_2\right)g^{\mu \nu}S^{\alpha 2}_{\alpha}]\nonumber
	\\
	&&-2\beta_2\mathbf{e}_{(12)}g^{\mu \nu}S^{\alpha}_{\alpha 1} +V_\nu^- \{ \sqrt{2}\left( \mathbf{e}_{[23]} -i\mathbf{e}_{[24]} \right)V^{\mu \nu +}+\sqrt{2}\big(\mathbf{e}_{(23)}+  
	\\
	&&-i\mathbf{e}_{(24)}\big)[ \beta_+S^{\mu \nu +}+\left( \beta_+ +17\rho_+ \right)g^{\mu \nu}S^{\alpha +}_\alpha  ] - \beta_2\left(\mathbf{e}_{(33)} + \mathbf{e}_{(44)}\right)\big( g^{\mu \nu}S^{\alpha +}_\alpha  \nonumber
	\\
	&&+2S^{\mu \nu +} \big) \}+ V_\nu^+ \{ \sqrt{2}\left( \mathbf{e}_{[23]} +i\mathbf{e}_{[24]} \right)V^{\mu \nu -}+\sqrt{2}\left(\mathbf{e}_{(23)} +i\mathbf{e}_{(24)}\right) \nonumber
	\\
	&&\left( \beta_- +17\rho_- \right)g^{\mu \nu}S^{\alpha -}_\alpha - \beta_2\left(\mathbf{e}_{(33)} + \mathbf{e}_{(44)}\right)\left( g^{\mu \nu}S^{\alpha -}_\alpha + 2S^{\mu \nu -} \right) \}\nonumber
\end{eqnarray}

Considering positive charged photon,

$V^{+}_{\mu}$ spin-1:
\begin{eqnarray}
	&&\partial_{\nu} \big\{2\left( a_3 + \beta_+\beta_-\right)V^{\nu \mu -} + 2b_3\left(\mathbf{e}^{[-1][\nu \mu]} + \mathbf{e}^{[-2][\nu \mu]}\right)\big\}+\nonumber
	\\
	 &&- \mathbf{m}_V^2 V^{\mu -}_{T} + l_{V T}^{\mu -} + c_{V T}^{\mu -} = j^{\mu -}_{V T}
\end{eqnarray}

$V^{+}_{\mu}$ spin-0:
\begin{equation}
	\partial_{\nu} \{ 4s_-g^{ \mu \nu}S^{\alpha -}_\alpha +c_- g^{\mu \nu}\left( \mathbf{e}^{(-1)\alpha}_\alpha + \mathbf{e}^{(-2)\alpha}_\alpha \right)   \} -\mathbf{m}_V^2 V^{\mu -}_L + l^{\mu -}_{V L} +  c^{\mu -}_{V L} = j^{\mu -}_{V L}
\end{equation}
where
 \begin{eqnarray}
 	&&l^{\mu -}_{V} = -V_\nu^{-}\big\{\big[ 34\left( \mathbf{e}_{(33)} + \mathbf{e}_{(44)} \right)g^{\mu \nu} \left( \mathbf{e}^{(11)\alpha}_\alpha + \mathbf{e}^{(22)\alpha}_\alpha \right) \big]\big\} \nonumber
 	\\
 	&&-V_\nu^+\big\{16\left(\mathbf{e}_{(33)} -\mathbf{e}_{(44)}  \right)g^{\mu \nu}\mathbf{e}^{(--)\alpha}_{\alpha} +
 	+32i\mathbf{e}_{(34)}\mathbf{e}_{(34)}\mathbf{e}^{(--34)\alpha}_{\alpha}\big\}
 \end{eqnarray}
\begin{eqnarray}
	&&c^{\mu -}_V = -A_{\nu}\{ \sqrt{2}\left( \mathbf{e}_{[13]} +i\mathbf{e}_{[14]} \right)\big(\mathbf{e}^{[-1][\mu \nu]}  +  \mathbf{e}^{[-2][\mu \nu]}\big) +  \nonumber
	\\
	&&\sqrt{2}\left(\mathbf{e}_{(13)} + i\mathbf{e}_{(14)} \right) (\mathbf{e}^{(-1)(\mu \nu)}+ g^{\mu \nu}\mathbf{e}^{(-1)\alpha}_\alpha + \mathbf{e}^{(-2)(\mu \nu)}\nonumber
	\\
	 &&+ g^{\mu \nu}\mathbf{e}^{(-2)\alpha}_\alpha )+ 16\sqrt{2}\left(\mathbf{e}_{(13)} + i\mathbf{e}_{(14)}\right)g^{\mu \nu}\big(  \mathbf{e}^{(-1)\alpha}_\alpha \nonumber
	\\
	&&+ \mathbf{e}^{(-2)\alpha}_\alpha \big)\}-U_\nu\{ \sqrt{2}\left( \mathbf{e}_{[23]} + i\mathbf{e}_{[24]}\right)\big(\mathbf{e}^{[-1][\mu \nu]} +  \mathbf{e}^{[-2][\mu \nu]} \big) \nonumber
	\\
	&& +  \sqrt{2}\left( \mathbf{e}_{(23)} + i\mathbf{e}_{(24)}\right)(\mathbf{e}^{(-1)(\mu \nu)}+ \mathbf{e}^{(-2)\alpha}_\alpha+g^{\mu \nu}\mathbf{e}^{(-1)\alpha}_\alpha  
	\\
	&&+ \mathbf{e}^{(-2)(\mu \nu)}  + g^{\mu \nu}\mathbf{e}^{(-2)\alpha}_\alpha ) + 16g^{\mu \nu} \mathbf{e}^{(-1)\alpha}_\alpha\}\nonumber
	\\
	&&-V_\nu^-\{ 4i\mathbf{e}_{[34]}\mathbf{e}^{[+-][\mu \nu]} +4i\mathbf{e}_{[34]}\mathbf{e}^{[12][\mu \nu]} + 2\big( \mathbf{e}^{(11)(\mu \nu)} + \mathbf{e}^{(-2)(\mu \nu)}   \nonumber
	\\
	&&+ g^{\mu \nu}\mathbf{e}^{(-2)\alpha}_\alpha ) + 16g^{\mu \nu} \mathbf{e}^{(-1)\alpha}_\alpha\}+2\left(\mathbf{e}_{(33)} + \mathbf{e}_{(44)}\right)\mathbf{e}^{(+-)(\mu \nu)}  +\nonumber
	\\
	&&+ 2\left(\mathbf{e}_{(33)} + \mathbf{e}_{(44)}\right)g^{\mu \nu}\mathbf{e}^{(+-)\alpha}_\alpha\} + V_\nu^+\{ \left( \mathbf{e}_{(33)} - \mathbf{e}_{(44)}\right)\mathbf{e}^{(--)(\alpha)}_\alpha \nonumber
	\\
	&&+ 2 i\mathbf{e}_{(34)}\mathbf{e}^{(--34)(\mu \nu)} \}\nonumber
\end{eqnarray}
The current term is
\begin{eqnarray}
	&&j_V^{\mu -} = V_\nu^-\big\{ -2\mathbf{e}_{[34]}\left(b_1 F^{\mu \nu} + b_2 U^{\mu \nu }\right) + \left(\mathbf{e}_{(33)} + \mathbf{e}_{(44)}\right)[ \beta_1S^{\mu \nu}_1  \nonumber
	\\
	&&+ \left(\beta_1 + 17\rho_1\right)g^{\mu \nu}S^{\alpha}_{\alpha 1}+\beta_2S^{\mu \nu}_2+\left(\beta_2 + 17\rho_2\right)g^{\mu \nu}S^{\alpha}_{\alpha 2} ]\nonumber
	\\
	&&  -\beta_+\big(\mathbf{e}_{(13)}-i\mathbf{e}_{(14)}\big)\big(\frac{1}{\sqrt{2}}g^{\mu \nu}S^{\alpha}_{\alpha 1} +\sqrt{2}S^{\mu \nu }_1\big) + \nonumber
	\\
	&&-\beta_+\left(\mathbf{e}_{(23)} -i\mathbf{e}_{(24)}\right)\big(\frac{1}{\sqrt{2}}g^{\mu \nu}S^{\alpha}_{\alpha 2} +\sqrt{2}S^{\mu \nu }_2\big)\big\} \nonumber
	\\
	&&+ A_\nu\{ \sqrt{2}\left(\mathbf{e}_{[13]} + i\mathbf{e}_{[14]} \right)V^{\mu \nu -} + \sqrt{2}\left( \mathbf{e}_{(13)} +i \mathbf{e}_{(14)}\right)[ \beta_-S^{\mu \nu -} 
	\\
	&&+\left(\beta_- +17\rho_-\right)g^{\mu \nu}S^{\alpha -}_\alpha ]- \beta_+\big(\mathbf{e}_{(13)} - i\mathbf{e}_{(14)} \big)\big(\frac{1}{\sqrt{2}}g^{\mu \nu}S^{\alpha -}_{\alpha} \nonumber
	\\
	&&+ \sqrt{2}S^{\mu \nu -}\big) \} + U_\nu\big\{ \sqrt{2}\left(\mathbf{e}_{[23]} + i\mathbf{e}_{[24]} \right)V^{\mu \nu -} +\nonumber
	\\
	&&+ \sqrt{2}\left( \mathbf{e}_{(23)} +i \mathbf{e}_{(24)}\right)\big[ \beta_-S^{\mu \nu -} + \left(\beta_- +17\rho_-\right)g^{\mu \nu}S^{\alpha -}_\alpha \big]\nonumber
	\\
	&&-\beta_+\left(\mathbf{e}_{(23)} - i\mathbf{e}_{(24)} \right)\big(1/\sqrt{2}g^{\mu \nu}S^{\alpha -}_{\alpha} + \sqrt{2}S^{\mu \nu -}\big) \big\}\nonumber
\end{eqnarray}

Considering negative charged photon,

$V^{-}_{\mu}$ spin-1:
\begin{eqnarray}
	&&\partial_{\nu} \{2\left( a_3 + \beta_+\beta_-\right)V^{\nu \mu +} + 2 b_3 \left(\mathbf{e}^{[+1][\nu \mu]} + \mathbf{e}^{[+2][\nu \mu]}\right)\} \nonumber
	\\
	&&- \mathbf{m}_V^2 V^{\mu +}_{T} + l_{V T}^{\mu +} + c_{V T}^{\mu +} = j^{\mu +}_{V T}
\end{eqnarray}

$V^{-}_{\mu}$ spin-0:
\begin{equation}
	\partial_{\nu} \{ 4s_+g^{ \mu \nu}S^{\alpha +}_\alpha +c_+ g^{\mu \nu}\left( \mathbf{e}^{(+1)\alpha}_\alpha + \mathbf{e}^{(+2)\alpha}_\alpha \right)   \} -\mathbf{m}_V^2 V^{\mu +}_L + l^{\mu +}_{V L} +  c^{\mu +}_{V L} = j^{\mu +}_{V L}
\end{equation}
\begin{eqnarray}
	&&l^{\mu + }_{V} = -V_\nu^+\{\left[ 34\left( \mathbf{e}_{(33)} + \mathbf{e}_{(44)} \right)g^{\mu \nu} \left( \mathbf{e}^{(11)\alpha}_\alpha + \mathbf{e}^{(22)\alpha}_\alpha \right) \right]\}\nonumber
	\\
	&& -V_\nu^-\{16\left(\mathbf{e}_{(33)} -\mathbf{e}_{(44)}  \right)g^{\mu \nu}\mathbf{e}^{(++)\alpha}_{\alpha}+
	+32i\mathbf{e}_{(34)}\mathbf{e}_{(34)}\mathbf{e}^{(++34)\alpha}_{\alpha}\}
\end{eqnarray} 
\begin{eqnarray}
	&&c^{\mu +}_V = -A_{\nu}\big\{ \sqrt{2}\left(\mathbf{e}_{[13]} -i\mathbf{e}_{[14]} \right)\big(\mathbf{e}^{[+1][\mu \nu]}  + \mathbf{e}^{[+2][\mu \nu]} \big)  \nonumber
	\\
	&&+ \sqrt{2}\big(\mathbf{e}_{(13)} - i\mathbf{e}_{(14)} \big)\big(\mathbf{e}^{(+1)(\mu \nu)} +
	g^{\mu \nu}\mathbf{e}^{(+1)\alpha}_\alpha + \mathbf{e}^{(+2)(\mu \nu)} \nonumber
	\\
	&&+ g^{\mu \nu}\mathbf{e}^{(+2)\alpha}_\alpha + 16g^{\mu \nu}  \mathbf{e}^{(+1)\alpha}_\alpha + 16g^{\mu \nu}\mathbf{e}^{(+2)\alpha}_\alpha \big)\big\}\nonumber
	\\
	&&-U_\nu\big\{ \sqrt{2}\left( \mathbf{e}_{[23]} - i\mathbf{e}_{[24]}\right)\big(\mathbf{e}^{[+1][\mu \nu]} + \mathbf{e}^{[+2][\mu \nu]}\big) +\nonumber
	\\
	&&+  \sqrt{2}\left( \mathbf{e}_{(23)} - i\mathbf{e}_{(24)}\right)\big(\mathbf{e}^{(+1)(\mu \nu)} +g^{\mu \nu}\mathbf{e}^{(+1)\alpha}_\alpha + \mathbf{e}^{(+2)(\mu \nu)}\nonumber  
	\\
	&&+ g^{\mu \nu}\mathbf{e}^{(+2)\alpha}_\alpha  + 16g^{\mu \nu}\mathbf{e}^{(+1)\alpha}_\alpha + 16g^{\mu \nu}\mathbf{e}^{(+2)\alpha}_\alpha \big) \big\}
	\\
	&&-V_\nu^+\{ 4i\mathbf{e}_{[34]}\mathbf{e}^{[+-][\mu \nu]} + 4i\mathbf{e}_{[34]}\mathbf{e}^{[12][\mu \nu]} + 2\big( \mathbf{e}^{(11)(\mu \nu)} \nonumber
	\\
	&&+ \mathbf{e}^{(22)(\mu \nu)} + \mathbf{e}^{(12)(\mu \nu)}+g^{\mu \nu}\mathbf{e}^{(12)\alpha}_\alpha\big)+2\left(\mathbf{e}_{(33)} + \mathbf{e}_{(44)}\right)\mathbf{e}^{(+-)(\mu \nu)}\nonumber
	\\
	 &&+ 2\left(\mathbf{e}_{(33)} + \mathbf{e}_{(44)}\right)g^{\mu \nu}\mathbf{e}^{(+-)\alpha}_\alpha\} + \nonumber
	\\
	&&+V_\nu^-\{ \left( \mathbf{e}_{(33)} - \mathbf{e}_{(44)}\right)\mathbf{e}^{(++)(\alpha)}_\alpha+2 i\mathbf{e}_{(34)}\mathbf{e}^{(++34)(\mu \nu)} \}\nonumber
\end{eqnarray}
The current term is
\begin{eqnarray}
	&&j_V^{\mu +} = V_\nu^+\big\{ -2\mathbf{e}_{[34]}\left(b_1 F^{\mu \nu} + b_2U^{\mu \nu }\right) + \left(\mathbf{e}_{(33)} + \mathbf{e}_{(44)}\right)\big[ \left(\beta_1 + 17\rho_1\right)g^{\mu \nu}S^{\alpha}_{\alpha 1}\nonumber
	\\
	&&\beta_1S^{\mu \nu}_1+ \beta_2S^{\mu \nu}_2 +\left(\beta_2 + 17\rho_2\right)g^{\mu \nu}S^{\alpha}_{\alpha 2} \big]  -\beta_-\left(\mathbf{e}_{(13)} +i\mathbf{e}_{(14)}\right)
	\big(1/\sqrt{2}g^{\mu \nu}S^{\alpha}_{\alpha 1} \nonumber
	\\
	&&+\sqrt{2}S^{\mu \nu }_1\big)  
	-\beta_-\left(\mathbf{e}_{(23)} +i\mathbf{e}_{(24)}\right)\big(\frac{1}{\sqrt{2}}g^{\mu \nu}S^{\alpha}_{\alpha 2} +\sqrt{2}S^{\mu \nu }_2\big)\big\}+ A_\nu\{ \sqrt{2}\big(\mathbf{e}_{[13]} +
	\\
	&&- i\mathbf{e}_{[14]} \big)V^{\mu \nu +} + \sqrt{2}\left( \mathbf{e}_{(13)} -i \mathbf{e}_{(14)}\right)[ \beta_+S^{\mu \nu +}+\left(\beta_+ +17\rho_+\right)g^{\mu \nu}S^{\alpha +}_\alpha ]+\nonumber
	\\
	&&-\beta_-\left(\mathbf{e}_{(13)} + i\mathbf{e}_{(14)} \right)\big(1/\sqrt{2}g^{\mu \nu}S^{\alpha +}_{\alpha} + \sqrt{2}S^{\mu \nu +}\big) \} + U_\nu\big\{ \sqrt{2}\big(\mathbf{e}_{[23]} +\nonumber
	\\
	&&-i\mathbf{e}_{[24]} \big)V^{\mu \nu +} + \sqrt{2}\left( \mathbf{e}_{(23)} -i \mathbf{e}_{(24)}\right)\big[ \beta_+S^{\mu \nu +} + \left(\beta_+ + 17\rho_+ \right)g^{\mu \nu}S^{\alpha +}_{\alpha} \big] \nonumber
	\\
	&&- \beta_-\left(\mathbf{e}_{(23)} + i\mathbf{e}_{(24)} \right)\big(\frac{1}{\sqrt{2}}g^{\mu \nu}S^{\alpha +}_{\alpha} + \sqrt{2}S^{\mu \nu +}\big) \big\}\nonumber
\end{eqnarray}
A fields-fields EM is encountered. EM energy is developed with features beyond $E = mc^2$ and electric charge interaction. New opportunities for electric-magnetic fields pairs, mass, electric charge, spin are expressed. The EM pairs develop relationships beyond Maxwell, for instance, collective EM pairs are created. Fields are attached to mass as interconvertible. The energy-mass relationship is expressed without Higgs. Fields converted to masses are obtained through London and conglomerates terms. The interactive mass appears to work as the source.

A whole interlaced dynamics is expressed. New properties from such fundamental whole dynamics laws are expected. First, with NLED. Physics contains critical electric and magnetic fields $E_{c}= \frac{m^2_{e}c^3}{e h} \sim 1.32 \times 10^{18} V\cdot m^-1$, $B_c = \frac{m^2_{e}c^2}{e \hbar} \sim 4.41 \times 10^9 T$ which non-linearities have to be understood theoretically. Consider a fundamental NLED instead of effective models. Second, a physics-based on correlated fields is derived. Faraday lines are obtained for fields strengths. Third, a duality similar Seilberg-Witten equations are expected [63], [49]. Fourth, it hypothesis different EM phases, as Mandelstam considered non-abelian theories [64]. Fifth, whole equations introduce the evolution scope.  A quanta evolution physics where physical entities are submitted to fields of environment, free will, and adaptation. The quadruplet fields family is expressing a field's environment. Quanta generated from fields set and with a chance. The whole quanta with a change are determined from the free coefficients at quadruplet fields set Lagrangian. Quanta free will appears from circumstances given by free coefficients. Quanta adaptation appears as physics depending on energy distribution, as the system has minimal energy. Another type of adaptation should be considered from mathematics. Through various possible fields, solutions are to be derived from a nonlinear system.

\section{Bianchi Identities}
The third type of equation, to analyze the system, is constituted by eqs. (\ref{Usual photon} -\ref{Charge photon}) are the Bianchi identities. The following granular expressions are derived
\begin{equation}
	\partial_\mu F_{\nu \rho} + \partial_\nu F_{\rho \mu } + \partial_\rho F_{\mu \nu} = 0
\end{equation}
\begin{equation}
	\partial_\mu U_{\nu \rho } + \partial_\nu U_{\rho \mu } + \partial_\rho U_{\mu \nu } = 0
\end{equation}
\begin{equation}
	\partial_\mu V^{\pm}_{\nu \rho } + \partial_\nu V^{\pm}_{\rho \mu } + \partial_\rho V^{\pm}_{\mu \nu } = 0
\end{equation}

They will be complemented by twelve more collective Bianchi identities corresponding to each physical collective field. For the collective antisymmetric tensor, one gets
\begin{eqnarray}
	&&\partial_\mu \mathbf{e}^{[12]}_{[\nu \rho]} + \partial_\nu \mathbf{e}^{[12]}_{[\rho \mu]} + \partial_\rho \mathbf{e}^{[12]}_{[\mu \nu]} = \mathbf{e}_{[12]}\{A_\mu U_{\nu \rho} + A_\nu U_{\rho \mu}\nonumber
	\\
	&& + A_\rho U_{\mu \nu} -U_\mu F_{\nu \rho} - U_\nu F_{\rho \mu} + U_\rho F_{\mu \nu}  \},
\end{eqnarray}
\begin{eqnarray}\label{Bianchi identity 6.5}
	&&\partial_\mu \mathbf{e}^{[+-]}_{[\nu \rho]} + \partial_\nu \mathbf{e}^{[+-]}_{[\rho \mu]} + \partial_\rho \mathbf{e}^{[+-]}_{[\mu \nu]} = -i\mathbf{e}_{[34]} \{ V_\mu^{+}V^{-}_{\nu \rho} + V_\nu^{+}V^{-}_{\rho \mu} +\nonumber
	\\
	&&+V_\rho^{+}V^{-}_{\mu \nu} -  V_\mu^{-}V^{+}_{\nu \rho} - V_\nu^{-}V^{+}_{\rho \mu} - V_\rho^{-}V^{+}_{\mu \nu} \},
\end{eqnarray}
\begin{eqnarray}
	&&\partial_\mu \left( \mathbf{e}^{[+1]}_{[\nu \rho]} + \mathbf{e}^{[-1]}_{[\nu \rho]} \right) + \partial_\nu \left( \mathbf{e}^{[+1]}_{[\rho \mu]} + \mathbf{e}^{[-1]}_{[\rho \mu]} \right)  + \partial_\rho \left( \mathbf{e}^{[+1]}_{[\mu \nu]} + \mathbf{e}^{[-1]}_{[\mu \nu]}  \right) =\nonumber
	\\
	&&=  \frac{1}{\sqrt{2}}\left( \mathbf{e}_{[13]} + i\mathbf{e}_{[14]} \right)\{ A_\mu V^{+}_{\nu \rho} + A_\nu V^{+}_{\rho \mu} + A_\rho V^{+}_{\mu \nu}  +V^{+}_\mu F_{\nu \rho} +\nonumber
	\\
	&&+   V^{+}_\nu F_{\rho \mu} +  V^{+}_\mu F_{\nu \rho}   \} +
	\frac{1}{\sqrt{2}}\left( \mathbf{e}_{[13]} + i\mathbf{e}_{[14]} \right)\{ A_\mu V^{-}_{\nu \rho} + A_\nu V^{-}_{\rho \mu} + 
	\\
	&&A_\rho V^{-}_{\mu \nu} + V^{-}_\mu F_{\nu \rho} +  V^{-}_\nu F_{\rho \mu} +  V^{-}_\mu F_{\nu \rho} \},\nonumber
\end{eqnarray}
and
\begin{eqnarray}\label{bianchi identity 6.7}
	&&\partial_\mu \left( \mathbf{e}^{[+2]}_{[\nu \rho]} + \mathbf{e}^{[-2]}_{[\nu \rho]} \right) + \partial_\nu \left( \mathbf{e}^{[+2]}_{[\rho \mu]} + \mathbf{e}^{[-2]}_{[\rho \mu]} \right)  + \partial_\rho \left( \mathbf{e}^{[+2]}_{[\mu \nu]} + \mathbf{e}^{[-2]}_{[\mu \nu]}  \right) =\nonumber
	\\
	&&=  \frac{1}{\sqrt{2}}\left( \mathbf{e}_{[23]} + i\mathbf{e}_{[24]} \right)\{ U_\mu V^{+}_{\nu \rho} + U_\nu V^{+}_{\rho \mu} + U_\rho V^{+}_{\mu \nu}  + V^{+}_\mu U_{\nu \rho}
	\\
	&& +  V^{+}_\nu U_{\rho \mu} +  V^{+}_\mu U_{\nu \rho}   \} +
	\frac{1}{\sqrt{2}}\left( \mathbf{e}_{[23]} + i\mathbf{e}_{[24]} \right)\{ U_\mu V^{-}_{\nu \rho} + U_\nu V^{-}_{\rho \mu} + \nonumber
	\\
	&&U_\rho V^{-}_{\mu \nu} + V^{-}_\mu U_{\nu \rho} +  V^{-}_\nu U_{\rho \mu} +  V^{-}_\mu U_{\nu \rho} \}.\nonumber
\end{eqnarray}

For collective symmetric tensor, it yields
\begin{equation}\label{bianchi identity 6.8}
	\partial_\mu \mathbf{e}^{(11)}_{(\nu \rho)} + \partial_\nu \mathbf{e}^{(11)}_{(\rho \mu)} + \partial_\rho \mathbf{e}^{(11)}_{(\mu \nu)} = \mathbf{e}_{(11)}\{ A_\mu S_{\nu \rho 1} +  A_\nu S_{\rho \mu 1} +A_\rho S_{\mu \nu 1} \},
\end{equation}
\begin{equation}
	\partial_\mu \mathbf{e}^{(22)}_{(\nu \rho)} + \partial_\nu \mathbf{e}^{(22)}_{(\rho \mu)} + \partial_\rho \mathbf{e}^{(22)}_{(\mu \nu)} = \mathbf{e}_{(22)}\{ U_\mu S_{\nu \rho 2} +  U_\nu S_{\rho \mu 2} + U_\rho S_{\mu \nu 2} \},
\end{equation}
\begin{equation}
	\partial_\mu \mathbf{e}^{(12)}_{(\nu \rho)} + \partial_\nu \mathbf{e}^{(12)}_{(\rho \mu)} + \partial_\rho \mathbf{e}^{(12)}_{(\mu \nu)} = \mathbf{e}_{(12)}\{ A_\mu S_{\nu \rho 2} +  A_\nu S_{\rho \mu 2} +A_\rho S_{\mu \nu 2} + U_\mu S_{\nu \rho 1} +  U_\nu S_{\rho \mu 1} + U_\rho S_{\mu \nu 1} \},
\end{equation}
\begin{eqnarray}
	&&\partial_\mu \mathbf{e}^{(+-)}_{(\nu \rho)} + \partial_\nu \mathbf{e}^{(+-)}_{(\rho \mu)} + \partial_\rho \mathbf{e}^{(+-)}_{(\mu \nu)} = \frac{1}{2}\left( \mathbf{e}_{(33)} + \mathbf{e}_{44}\right)\{ V^{-}_\mu S^{+}_{\nu \rho } +  V^{-}_\nu S^{+}_{\rho \mu} \nonumber
	\\
	&&+V^{-}_\rho S^{+}_{\mu \nu } + V^{+}_\mu S^{-}_{\nu \rho } +  V^{+}_\nu S^{-}_{\rho \mu} +V^{+}_\rho S^{-}_{\mu \nu } \},
\end{eqnarray}
\begin{eqnarray}
	&&\partial_\mu \left( \mathbf{e}^{(+1)}_{(\nu \rho)} + \mathbf{e}^{(-1)}_{(\nu \rho)} \right) + \partial_\nu \left( \mathbf{e}^{(+1)}_{(\rho \mu)} + \mathbf{e}^{(-1)}_{(\rho \mu)} \right)  + \partial_\rho \left( \mathbf{e}^{(+1)}_{(\mu \nu)} + \mathbf{e}^{(-1)}_{(\mu \nu)}  \right) =  \nonumber
	\\
	&&=\frac{1}{\sqrt{2}}\left( \mathbf{e}_{(13)} + i\mathbf{e}_{(14)}\right)\{ A_\mu S^{+}_{\nu \rho} + A_\nu S^{+}_{\rho \mu} + A_\rho S^{+}_{\mu \nu} + V^{+}_{\mu} S_{\nu \rho 1} + \nonumber
	\\
	&&V^{+}_{\nu} S_{\rho \mu 1} + V^{+}_{\rho} S_{\mu \nu 1} \} + \frac{1}{\sqrt{2}}\left( \mathbf{e}_{(13)} - i\mathbf{e}_{(14)}\right)\{ A_\mu S^{-}_{\nu \rho} + A_\nu S^{-}_{\rho \mu} 
	\\
	&&+ A_\rho S^{-}_{\mu \nu} + V^{-}_{\mu} S_{\nu \rho 1} + V^{-}_{\nu} S_{\rho \mu 1} + V^{-}_{\rho} S_{\mu \nu 1} \},\nonumber
\end{eqnarray} 
\begin{eqnarray}
	&&\partial_\mu \left( \mathbf{e}^{(+2)}_{(\nu \rho)} + \mathbf{e}^{(-2)}_{(\nu \rho)} \right) + \partial_\nu \left( \mathbf{e}^{(+2)}_{(\rho \mu)} + \mathbf{e}^{(-2)}_{(\rho \mu)} \right)  + \partial_\rho \left( \mathbf{e}^{(+2)}_{(\mu \nu)} + \mathbf{e}^{(-2)}_{(\mu \nu)}  \right) =\nonumber
	\\
	&&=\frac{1}{\sqrt{2}}\left( \mathbf{e}_{(23)} + i\mathbf{e}_{(24)}\right)\{ U_\mu S^{+}_{\nu \rho} + U_\nu S^{+}_{\rho \mu} + U_\rho S^{+}_{\mu \nu} + V^{+}_{\mu} S_{\nu \rho 2} + V^{+}_{\nu} S_{\rho \mu 2} \nonumber
	\\
	&&+ V^{+}_{\rho} S_{\mu \nu 2} \} + \frac{1}{\sqrt{2}}\left( \mathbf{e}_{(23)} - i\mathbf{e}_{(24)}\right)\{ U_\mu S^{-}_{\nu \rho} + U_\nu S^{-}_{\rho \mu} 
	+ U_\rho S^{-}_{\mu \nu} +
	\\
	&&+V^{-}_{\mu} S_{\nu \rho 2} + V^{-}_{\nu} S_{\rho \mu 2} +
	+ V^{-}_{\rho} S_{\mu \nu 2} \},\nonumber
\end{eqnarray}
\begin{eqnarray}
	&&\partial_\mu \left( \mathbf{e}^{(++)}_{(\nu \rho)} + \mathbf{e}^{(--)}_{(\nu \rho)} \right) + \partial_\nu \left( \mathbf{e}^{(++)}_{(\rho \mu)} + \mathbf{e}^{(--)}_{(\rho \mu)} \right)  + \partial_\rho \left( \mathbf{e}^{(++)}_{(\mu \nu)} + \mathbf{e}^{(--)}_{(\mu \nu)}  \right) =\nonumber
	\\
	&&=\frac{1}{2}\left(\mathbf{e}_{(33)} - \mathbf{e}_{(44)}\right)\{V^{+}_\mu V^{+}_{\nu \rho} +  V^{+}_\nu V^{+}_{\rho \mu} + V^{+}_\rho V^{+}_{\mu \nu} + 
	\\
	&&V^{-}_\mu V^{-}_{\nu \rho} +  V^{-}_\nu V^{-}_{\rho \mu} + V^{-}_\rho V^{-}_{\mu \nu}\},\nonumber
\end{eqnarray}
and
\begin{eqnarray}\label{bianchi identity 6.15}
	&&\partial_\mu \left( \mathbf{e}^{(++34)}_{(\nu \rho)} + \mathbf{e}^{(--34)}_{(\nu \rho)} \right) + \partial_\nu \left( \mathbf{e}^{(++34)}_{(\rho \mu)} + \mathbf{e}^{(--34)}_{(\rho \mu)} \right)  + \nonumber
	\\
	&&\partial_\rho \left( \mathbf{e}^{(++34)}_{(\mu \nu)} + \mathbf{e}^{(--34)}_{(\mu \nu)}  \right) = i\mathbf{e}_{(34)}\{V^{+}_\mu V^{+}_{\nu \rho} +  V^{+}_\nu V^{+}_{\rho \mu} + 
	\\
	&&V^{+}_\rho V^{+}_{\mu \nu} - V^{-}_\mu V^{-}_{\nu \rho} -  V^{-}_\nu V^{-}_{\rho \mu} - V^{-}_\rho V^{-}_{\mu \nu}\}.\nonumber
\end{eqnarray}
The above equations are showing induced relationships derived from the quadruplet correlations. New Faraday-Lenz-Neumann laws are developed [65]. It produces (4+4) equations of motion and 15 Bianchi identities.
Monopoles appear related to fields associations. The collective antisymmetric Bianchi identities, eqs. (\ref{Bianchi identity 6.5} - \ref{bianchi identity 6.7}) can be rewritten as
\begin{equation}
	\partial_\mu \tilde{e}^{\mu \nu} = j^{\nu}_B, \ \partial_{\nu}j^{\nu}_B=0 \label{Conservação da corrente elétrica}
\end{equation}
Eqs. (\ref{bianchi identity 6.8}-\ref{bianchi identity 6.15}) are also introducing sources.

Monopoles have a rich story in physics. The magnet slicing experiment was first performed in 1269 by the French scholar Peregrinus [66]. In 1931 Dirac introduced the magnetic charge [67]. Different monopoles scenarios follow as monopoles as gauge particles [68], GUTs [69], Kaluza-Klein [70], Superstrings [71], Quantum Gravity and Cosmology [72], Electroweak monopoles [73], monopoles produced in intense magnetic fields [74], monopoles and photons decay [75], monopoles condensated in superconductors' [76], monopoles and dark matters [77].

The four bosons EM introduces monopoles as potential fields conglomerates. Similar to 't Hooft and Polyakov monopoles based on Higgs fields [78]. Monopoles depending on fields contain opportunities at the condensed matter as spin ice monopoles [79].

\section{Invariants}
The extended Yang-Mills functional is invariant by the gauge group action [80]. Consequently, quadruplet invariants are derived. Lorentz's scalars and conserved charges are obtained. EM invariants are obtained under Lorentz transformations [81]. Diverse scalars are build up through combinations between fields strengths as $F_{\mu \nu}^2, F_{\mu \nu}U^{\mu \nu}, F_{\mu \nu}\mathbf{e}^{\mu \nu}$ and so on. 

Four types of fields conserved charges are provided. First, by considering the (4+4) transverse and longitudinal whole equations. 

For spin-1,
\begin{equation}
	\partial \cdot \left(\mathbf{m}^2 A^{T}_I + l^{T}_I + c^{T}_I - j^{T}_I\right) = 0
\end{equation}
and for spin-0,
\begin{equation}
	\Box\left(S^{\alpha}_{\alpha I} + \mathbf{e}_{\alpha I}^{\alpha}\right) = \partial \cdot \left(j - M - l - c\right)
\end{equation}

A second conservation law comes from antisymmetric collective Bianchi identities. From eq. (\ref{Conservação da corrente elétrica}), one gets
\begin{equation}\label{equação 7.3}
	\partial_{\mu}J^{\mu}_B = 0  \text{ where the source is }  J_{\mu}^{B} = \epsilon_{\mu \nu \rho \sigma} \mathbf{e}_{[IJ]}A_I^{\nu} A_J^{\rho \sigma}
\end{equation}
where eq. (\ref{equação 7.3}) is showing the presence of fields monopoles 

Noether current expresses a third continuity equation. A relevant aspect is that electric charge is associated with a group of fields not necessarily charged. The fourth conservation law is given by the second type of global conservation associated to the improved energy-momentum tensor, $\partial_\mu \theta^{\mu \nu} = 0$ [82]. The quadruplet EM flux develops conservation laws between energy dimity, Poynting vector, and stress tensor. They are expressed as
\begin{equation}
	\frac{\partial U}{\partial t} + \vec{\bigtriangledown}\cdot\vec{S} = 0 
\end{equation}
\begin{equation}
	\frac{\partial S_i}{\partial t} + \partial_j T_{j i} = 0
\end{equation}

A physicality beyond matter appears. The quadruplet is expressing physics, providing the meaning of the field's set. Fields conservation laws based on fields and fields relationships are obtained. 

\section{Nonlinear wave equations}

Maxwell focused on charge conservation and distribution; the four bosons electromagnetism introduces the charge transfer. A difference appears while Hertz's experiments in 1888 comprised Maxwell linear EM waves [83], a further investigation must be done. Nonlinear wave equations have been explored [84]. New types of transverse and longitudinal waves are obtained.

For $A_{\mu T}$,
\begin{equation}
	\Box\{ 4\left(a_1 + \beta_1 +a_1k_1\right)F^{\nu \mu} + 2b_1[\mathbf{e}^{[12][\nu \mu]} + (1+k_1)\mathbf{e}^{[+-][\nu \mu]}] \} = \partial^{\nu}d^{\mu}_{ A T} - \partial^{\mu}d^{\nu}_{ A T} 
\end{equation}
where
\begin{equation}
	d^{\mu}_{A T} = M^{A \mu}_{I T}+ j^{\mu}_{A T} - J^{\mu}_{N T} - k_2 J^{\mu}_{U T} - l_{A T}^{\mu} + c^{\mu}_{A T}
\end{equation}

For $A_{\mu L}$,
\begin{eqnarray}
	&&\Box\{s_{11}S^{\alpha}_{\alpha 1} + 2\left(\mathbf{e}^{(11)\alpha}_\alpha + \mathbf{e}^{(22)\alpha}_\alpha + \mathbf{e}^{(12)\alpha}_\alpha + \mathbf{e}^{(+-)\alpha}_\alpha \right)\} = \nonumber
	\\
	&&\partial \cdot\left(j_{A L} + t_{11}M^{A}_{g L} + \frac{1}{2}t_{11}J_{U L} -l_{A L} - c_{A L}\right) 
\end{eqnarray}

For $U_{\mu T}$,
\begin{eqnarray}
	&&\Box\{ 4\left(a_2 + \beta_2 +a_2k_2\right)U^{\nu \mu} + 2b_2[\mathbf{e}^{[12][\nu \mu]} + (1+k_2)\mathbf{e}^{[+-][\nu \mu]}] \} = \nonumber
	\\
	&&=\partial^{\nu}d^{\mu}_{U T} - \partial^{\mu}d^{\nu}_{U T} 
\end{eqnarray}
where
\begin{equation}
	d^{\mu}_{U T} = M^{U \mu}_{g T}+ j^{\mu}_{U T} - J^{\mu}_{N T} - k_1 J^{\mu}_{A T} -M^{\mu}_{U T} - l_{A T}^{\mu} + c^{\mu}_{A T}
\end{equation}

For $U_{\mu L}$,
\begin{eqnarray}
	&&\Box\{s_{22}S^{\alpha}_{\alpha 2} + 2\left(\mathbf{e}^{(11)\alpha}_\alpha + \mathbf{e}^{(22)\alpha}_\alpha + \mathbf{e}^{(12)\alpha}_\alpha + \mathbf{e}^{(+-)\alpha}_\alpha \right)\} = \nonumber
	\\
	&&=\partial \cdot\left(M^{U}_{g L} + j_{U L} + t_{22}J_{A L}- M_{U L} - l_{U L} - c_{U L}\right) 
\end{eqnarray}

For $V^-_{\mu T}$,
\begin{equation}
	\Box\{2\left(a_3 + \beta_+\beta_-\right)V^{\nu \mu + } + 2b_3\left(\mathbf{e}^{[+1][\nu \mu]} + \mathbf{e}^{[+2][\nu \mu]}\right)\} = \partial^{\nu}d^{\mu +}_{V T} - \partial^{\mu}d^{\nu +}_{V T} 
\end{equation}
where
\begin{equation}
	d^{\mu +}_V = j^{\mu +}_{V T} - M^{\mu +}_{V T} - l^{\mu +}_{V T} - c^{\mu +}_{V T}
\end{equation}

For $V^-_{\mu L}$,
\begin{equation}
	\Box \{4s_+S^{\alpha +}_{\alpha} + c_+\left(\mathbf{e}^{(+1)\alpha}_\alpha + \mathbf{e}^{(+2)\alpha}_\alpha \right)\} = \partial \cdot \left(j^{+}_{V L} - M^+_{V L} - l^{+}_{V L} - c^{+}_{V L}\right)
\end{equation}
For $V^+_{\mu T}$,
\begin{equation}
	\Box\{2\left(a_3 + \beta_+\beta_-\right)V^{\nu \mu -} + 2b_3\left(\mathbf{e}^{[-1][\nu \mu]} + \mathbf{e}^{[-2][\nu \mu]}\right)\} = \partial^{\nu}d^{\mu -}_{V T} - \partial^{\mu}d^{\nu -}_{V T} 
\end{equation}
where
\begin{equation}
	d^{\mu -}_V = j^{\mu -}_{V T} - M^{\mu -}_{V T} - l^{\mu -}_{V T} - c^{\mu -}_{V T}
\end{equation}
For $V^+_{\mu L}$,
\begin{equation}
	\Box \{4s_-S^{\alpha -}_{\alpha} + c_-\left(\mathbf{e}^{(-1)\alpha}_\alpha + \mathbf{e}^{(-2)\alpha}_\alpha \right)\} = \partial \cdot \left(j^{-}_{V L} - M^-_{V L} - l^{-}_{V L} - c^{-}_{V L}\right)
\end{equation}

\section{Whole Lorentz forces}

Special relativity defines force through the equation $\partial_{\mu}\theta^{\mu \nu} = f^{\nu}$ which means a change of energy and momentum between two systems made of fields and external sources [85]. This leads to us following Lagrangian density.
\begin{equation}
	L = L_{0} + A^{\mu I}j_{\mu I}
\end{equation}
where $j_{\mu I}$ means N-external currents, not necessarily corresponding to electric charge. The generic whole fields equation becomes

\begin{equation}
	\partial_\mu \left( F^{\nu \mu}_I + \mathbf{e}_I^{[\nu \mu]} + g^{\nu \mu}S_{\alpha I}^{\alpha} + \mathbf{e}_{I}^{(\nu \mu)}\right) + \frac{1}{2}\mathbf{m}^2_{I}A^{\mu}_I = J^{\mu}_I(A) + j^{\mu}_I
\end{equation}

It yields the fundamental field's equation [49]
\begin{equation}
	f^{\nu} = f^{\nu}_L + f^{\nu}_M + f^{\nu}_E
\end{equation}
where
\begin{eqnarray}
	f^{\nu}_L = 4F^{\nu \mu}_I j^{I}_{\mu}, & f^{\nu}_M = -2\mathbf{m}^2_I A^{\mu}_I(\partial^{\mu}A^{I}_{\mu}), & f^{\nu}_E = 4A^{\nu}_I\left(\partial^{\mu}J_{\mu}^I\left(A\right)\right)
\end{eqnarray}
where $A_{\mu I} \equiv \{A_{\mu}, U_{\mu}, V_{\mu}^{\pm}\}$ and $J_{\mu I}$ the corresponding fields current. $f^{\nu}_L$ means the extended Lorentz force, $f^{\nu}_M$ a mass force and $f^{\nu}_E$ a fields environmental force.

\section{Conclusion}

EM under charge transference is studied. A new border is set up. Three historical moments retreat the EM development. The third EM period is crossed, it is between fields and fields. A new physics for Faraday lines of force. The meaning of electric and magnetic fields is expanded. New terms, are provided EM fields working as own sources, fields and matter.

A four bosons electromagnetism appears. The charges set $\{+,0,-\}$ intermediates a quadruplet $\{A_{\mu},U_{\mu},V^{\pm}_{\mu}\}$. A whole abelian quantum field theory is derived under electric charge symmetry.  New properties for the EM flow are determined. Physicalities from nonlinear abelian fields, new EM regimes, and electric charge symmetry.

A physics beyond Maxwell is opened. The charge transfer phenomenology is explored. Nonlinear Faraday lines of force appear. Potential fields are empowered by their own charges. The four bosons EM moves electric charge symmetry from $U(1)$ to $U(1) \times SO(2)_{global}$. A neutral EM energy appears. The third nature for electric charge is included. Zero electric charges is introduced as an abelian quantum number.

Nevertheless, a model should first be tested on how far it covers old limitations before proposing new properties. Four consistencies are observed from the electromagnetic whole equations. First, a nonlinear EM is obtained at a fundamental level. Second, potential fields are coupled with granular and collective fields' strengths instead of subsidiaries. As in quantum mechanics, appear explicitly at equations. A result supporting the 1845 Faraday effect in 1845 between light and magnetism [2] and Aharanov-Bohm and Aharanov-Casher experiments [86]. Third, the polarization and magnetization vectors are derived from the first principles [57]. Identified in the equations of motion with the collective fields strengths. Fourth, light is no more a passive consequence of electric charge oscillations. The photon equation of motion, eq. (\ref{Effective A}), is showing a photon field with its charge depending on fields and with self-interacting photons at tree level.

Based on such consistencies, new EM features are expressed. EM phenomena are accomplished by four photons. Nonlinear whole relativistic equations with new relationships between electric and magnetic. Granular and collective fields' strengths share a common dynamic. Although under the same Lorentz symmetry, the spin-1 and spin-0 sectors are decoupled. Two spin dynamics succeed. Each sector contains a proper physics. Their quanta differ on spin, masses, charges, coupling constants.

Thus, three observances are supporting this search for a new EM energy. Two electric charge phenomenology, four surpassed Maxwell limitations, and a renormalizable theory [87]. These achievements bring confidence to the model being investigated. Explore their fields, energy, and electric charge new significances.

The first meaning is on physical features depending just on fields. A deeper relationship between energy and matter is performed. Beyond $E = mc^2$, fields and matter physics are developed. London and conglomerate expressions at eqs. (\ref{London terms}) and (\ref{conglomerates terms}) are introducing a relationship between fields and mass. Fields become physical entities with their own character. Phenomena as superconductivity [55], dark mass [88], dark energy [89] may have consequences for fields and fields of physics.  Analysing the energy-momentum tensor, the quadruplet model also includes fields dependences on energy, momentum, stress tensor [82]. 

The second meaning is that six new interlaced EM regimes are produced. They are systemic, nonlinear, neutral, spintronics, photonics, electroweak. First, a field's set of physics is constituted. While Maxwell related just the pair $\{\vec{\mathbf{E}}, \vec{\mathbf{B}}\}$  under four equations, eqs. (\ref{Usual photon}-\ref{Charge photon}) expands to a system expressed by a set of potential fields associations [43]. A whole EM is determined under a set causality. The quadruplet integrates twenty-three equations. Their variables and parameters are expressed in the function of the whole. The whole Lorentz force is derived.

The second regime is nonlinearity. There are sixteen nonlinear EM models in literature [36]. Most of them are effective theories derived from Born-Infeld and Euler-Heisenberg in the 1930s. Diversely, eqs. (\ref{Usual photon}-\ref{Charge photon}) propose a fundamental model. Instead of powering, as $\vec{\mathbf{E}}^2$, introduce as sources interactions between fields strengths and potentials, as $\vec{\mathbf{E}}\cdot\vec{A}$, with adimensional coupling constants. A renormalizable and unitary nonlinear EM is produced.

A neutral EM follows. EM energy is more primitive than electric charge. The interconnected charges set $\{+,0,-\}$ introduces the neutral charge physicality. An EM energy with a coupling beyond Maxwell-QED. Potential fields, electric and magnetic fields developing a physics not restricted to the fine structure constant. The corresponding equations of motion, energy-momentum tensor, and forces are showing sectors with couplings diverse from electric charge. The zero charge from passive becomes active.

The fourth EM regime is on spintronics. Originally, spin was introduced heuristically, through the magnetic moment interaction $\vec{\mu}\cdot\vec{\mathbf{B}}$ [90]. The four bosons model introduces spin from first principles as Lie Algebra valued [61]. Defining the potential field $A_{\mu, \kappa \lambda}$ where the first indices mean the space-time symmetry, $x'_{\mu} = \Lambda_{\mu}^{\nu}x_{\nu}$, and others the $SO(1,3)$ rotation, $A'_{\mu} = \left(e^{i\omega_{\alpha \beta}\Sigma^{\alpha \beta}}\right)_{\mu}^{\nu}A_{\nu}$, one incorporates as ab initio the spin generator $\left( \Sigma_{\alpha \beta}\right)_{\mu}^{\nu}$. Similarly to Yang-Mills, $A_{\mu} \equiv A_{\mu a}t_a$, the quadruplet fields are introduced as spin algebra value. Consequently, the electric and magnetic dipole couplings are derived from first principles through spin fields interactions, as $gF^{\alpha \beta}\left(\Sigma_{\alpha \beta}\right)^{\nu}_\mu A_\nu$. 

A fifth EM regime is photonics. Consider the photon as originally. Diversely from Big-Bang theory and Maxwell introduce a primordial light. Big-Bang derives the photon after the second phase transition $SU_{C}(3) \times SU_{L}(2) \times U_{Y}(1) \stackrel{SSB}{\rightarrow} SU_{C}(3)\times U_{em}(1)$ at first $10^{-10}s$ due to the Higgs presence at $10^{15}K$ temperature, $250GeV$ energy, $10^{19}T$ magnetic field [91]. Consequently, the Standard Model photon appears after spontaneous breaking symmetry, as $U_{em}(1)_{photon} = cos\theta_w \cdot U_{Y}(1) + sin \theta_w I_3$. Concurrently, Maxwell photon is a radiation from oscillating electric charge. It does not produce even its own electromagnetic fields.

The photon should be considered as cause and not effect. An original photon containing light invariance, ubiquitous lux, and self-interacting photons is expected from a new EM model. Consider the light emulation without any premise. Eq. (\ref{Effective A}) is introducing a physics with self-interacting photons not depending on electric charge. A nonlinearity to create matter from pure energy. A possibility to smash two photons and create particles.

The photonics turning point is on photon interaction. A discussion returning to 1934 with Delbruck and Breit-Wheeler [60]. In 1950 the $\gamma \gamma$ scattering was studied by Karplus and Neuman through QED [92]. Complemented in the decade by Skobov and Minguzzi [93]. Evidence for light by light scattering in heay-ion collisions is being explored at ATLAS [94]. The confirmation happened in 2017, when LHC measured the photon-photon interaction [95]. However, through one loop correction. Something is missing. Primordial photon should be manifested at fundamental physics. Scatterings $\gamma \gamma \to \gamma \gamma$ and $\gamma \gamma \to f \bar{f}$ at tree level. 

Astrophysics [96], photonics process in intense fields [97], nonlinear optics [98] and lasers [99] are studying on light non-linearity. Nonlinear photons may also be derived from intense EM fields as in pulsars ($10^8 T$), magnetars ($10^{11}T$). Photonic processes under intense EM fields are making physics return to later Delbruck scattering 85 years later [100]. Different propositions are being studied as the vacuum birefringence [101]. Lasers experiment in Padova [102], the Photon Collider in Germany (Vaccum Hohlraum), Star (BNL), Shangai Super-Laser (SULF) [103] and other lasers are being developed around the world. In 2026 the Xcels in Russia expect to look for the primordial photon.

A new EM formulation with the electron being substituted by the photon is expected. Photonics be constituted. Eq. (\ref{Effective A}) considers nonlinear photon field carrying its own charge, producing proper electric and magnetic fields and self-interactions without the electric charge presence. A fundamental NLED candidate to provide theoretical support to the deflection of light by light [104], quantum computer [105], two and three photons bound states, light molecules [106], liquid photons [107].

The last EM regime is electroweak. Interpret weak interaction as an EM sector. Photon, $Z_0, W^{\pm}$ be included together without asking for a Standard Model. Consider just charge transference physics and identify the corresponding four bosons as $\gamma, Z^0,W^{\pm}$. Recover the Abelian Standard Model version $U(1) \times SU(2) \stackrel{SSB}{\rightarrow} U_{em}$ through $U(1) \times SO(2)_{global}$. However, instead of finalizing with $U_{em}(1)$ symmetry, as SM does, the four bosons EM stars with abelian symmetry since the beginning. Take as guidance principle a modified electric charge conservation. And so, as an alternative to the Standard Model, the four bosons electromagnetism introduces a kinetic longitudinal sector with scalar photons [108],  mass without Higgs [52], photon including small mass [109], interaction $\gamma- Z_0$, anomalous triple gauge bosons coupling [110], spin (g-2) terms $F_{\mu \nu}Z^{\mu}Z^{\nu}$, $Z_{\mu \nu}A^{\mu} A^{\nu}$ and self-interacting photon at tree level. New scales of intensities and range are derived by intermediate fields through coupled nonlinear equations, mass terms, and coupling constants beyond electric charge [111]. 

The third EM meaning from charge transfer is on electric charge physics. Electromagnetism is being considered as anything that acts on an electric charge. However, it should primordially be a consequence of electric charge symmetry. The Symmetry $U(1) \times SO(2)_{global}$ moves the Maxwell charge physics from interactions and conservation law to embracing a new performance. It shows that electric charge physics is more primitive than being the source for electric-magnetic fields or acting as a coupling constant. Electric charge originally function is associative. Its symmetry physics is to integrate fields. As a consequence, the EM energy is derived from electric charge symmetry, without necessarily depending on its presence. 

This discrepancy between electric charge and EM energy is a fact already constituted in Maxwell. There, the energy flux at the electric circuit does not follow in wires. Yet, Faraday law does not depend on electric charge. Therefore, charge transfer extends but does not modify Maxwell's structure. An EM energy flux based on fields is processed. A field's physicality preceding matter is derived. Each potential field, $A_{\mu I}$, provides its own continuity flowing not directly related to electric charge. It yields physical laws based on electric charge as a quantum number but beyond electric charge as coupling constant.

A perspective for EM energy based on a general electric charge physics is envisaged. Electric charge symmetry is the prior definition for what EM is. The charge transfer symmetry rules new sectors and states an EM energy without the electric charge presence. Maxwell is rewritten as part of enlarged electromagnetism. A physics to be explored just based on fields. Faraday lines of force antecede the presence of physical constants as permissively and permeability.  There is a third electromagnetic historical period to be conquested.

\section*{Acknowledgment}
Luís Santiago Mendes would like to express his gratitute to FAPERJ (Rio de Janeiro state foundation) for supporting with an innovation scholarship.

%

\begin{appendices}
	\section{Appendix. Fields strengths gauge symmetry}
	The minimum condition for an observable to be considered is gauge invariance. The novelty when different families are rotating in the same group is that various coefficients are incorporated in theory. They are free parameters that can take any value without breaking the gauge symmetry. Consequently, it yields properties based on free coefficients. Freedom for gauge-invariant observables is constructed.
	
	Thus, the model derived from eqs(\ref{Lagrangiana}) is constituted by free coefficients. These adjustable parameters are a consequence of the symmetry abundance coming from the whole principle. It connects potential fields under free coefficients. From [49] the total number of free coefficients in $1 + (N-1)[N^3 +2N -4]$ where N is the number of potential fields under the same U(1) group. The eq. (\ref{Usual photon}) Lagrangian receives 205 free coefficients. A number is enough to adjust the field's strengths without losing gauge invariance [81]. Prescribe granular and collective fields strengths with gauge invariant properties. 
	
	There are two methods. Consider the Lagrangian gauge invariance [112] or separate the field's strengths separately, as in section 2. In this appendix, one relates the free coefficients relationships for establishing the field's strengths as a gauge invariant. For the granular symmetric, make 
	\begin{align}
		S_{\mu \nu 1}'= S_{\mu \nu 1} ;\  S_{\mu \nu 2}'= S_{\mu \nu 2}
	\end{align}
	under the conditions
	\begin{eqnarray}
		\beta_i\Omega_1^i\Omega^{-1}_{11} &=0, & \beta_i\Omega_2^i\Omega^{-1}_{21}= 0
	\end{eqnarray}
	and
	\begin{equation}
		S_{\mu \nu }^{\pm'}=e^{\pm i q \alpha} S_{\mu \nu}^{\pm}
	\end{equation}
	with
	\begin{align}
		\beta_i \Omega_3^i\Omega_{31}^{-1} &= 0, &  \beta_i \Omega_3^i\Omega_{41}^{-1} &= 0,&
		\beta_i \Omega_4^i\Omega_{31}^{-1} &=0, & \beta_i \Omega_4^i\Omega_{41}^{-1}&= 0
	\end{align}
	where $\Omega$ matrix is calculated in [51].
	
	For spin-0 case,
	\begin{align}
		S_{\alpha 1 }^{\alpha '} =  S_{\alpha 1 }^{\alpha} ;&  S_{\alpha 2}^{\alpha'}= S_{\alpha 2 }^{\alpha }
	\end{align}
	under
	\begin{equation}
		\rho_i\Omega_1^i\Omega_{11}^{-1} + \rho_i\Omega_2^i\Omega_{21}^{-1} = 0
	\end{equation}
	For the case,
	\begin{equation}
		S_{\alpha }^{\alpha \pm'}= e^{\pm i q \alpha} S_{\alpha}^{\alpha \pm}
	\end{equation}
	one gets
	\begin{align}
		\rho_i \Omega_3^i\Omega_{31}^{-1} &= 0, &  \rho_i \Omega_3^i\Omega_{41}^{-1} &= 0,&
		\rho_i \Omega_4^i\Omega_{31}^{-1} &=0, & \rho_i \Omega_4^i\Omega_{41}^{-1}&= 0
	\end{align}
	
	For antisymmetric collective fields strengths.
	\begin{eqnarray}
		\mathbf{e}_{[\mu \nu]}^{[12]'} = \mathbf{e}_{[\mu \nu]}^{[12]}; & \mathbf{e}_{[\mu \nu]}^{[+-]'} = \mathbf{e}_{[\mu \nu]}^{[+-]}\\
		\mathbf{e}^{[+1]'}_{[\mu \nu]} + \mathbf{e}^{[-1]'}_{[\mu \nu]'} &= e^{iq\alpha}\mathbf{e}^{[+1]}_{[\mu \nu]} +e^{-iq \alpha}\mathbf{e}^{[-1]}_{[\mu \nu]}\\
		\mathbf{e}^{[+2]'}_{[\mu \nu]} + \mathbf{e}^{[-2]'}_{[\mu \nu]} &= e^{iq\alpha}\mathbf{e}^{[+2]}_{[\mu \nu]} + e^{-iq \alpha}\mathbf{e}^{[-2]}_{[\mu \nu]}
	\end{eqnarray}
	it requires,
	\begin{align}
		\mathbf{e}_{[12]}\Omega_{11}^{-1} &= 0 ;& \mathbf{e}_{[12]}\Omega_{21}^{-1} &=0; &
		\mathbf{e}_{[34]}\Omega_{31}^{-1} &=0;& \mathbf{e}_{[34]}\Omega_{41}^{-1} &=0 \\
		\mathbf{e}_{[13]}\Omega_{31}^{-1} &=0 ;& \mathbf{e}_{[13]}\Omega_{41}^{-1} &=0; &
		\mathbf{e}_{[14]}\Omega_{31}^{-1} &=0 ;& \mathbf{e}_{[14]}\Omega_{41}^{-1} &=0\\
		\mathbf{e}_{[13]}\Omega_{11}^{-1} &=0 ;& \mathbf{e}_{[14]}\Omega_{11}^{-1} &=0;&
		\mathbf{e}_{[23]}\Omega_{31}^{-1} &=0 ;& \mathbf{e}_{[23]}\Omega_{41}^{-1} &=0\\
		\mathbf{e}_{[24]}\Omega_{31}^{-1} &=0 ;& \mathbf{e}_{[24]}\Omega_{41}^{-1} &=0;&
		\mathbf{e}_{[23]}\Omega_{11}^{-1} &=0 ;& \mathbf{e}_{[24]}\Omega_{11}^{-1} &=0
	\end{align}
	
	For symmetric collective fields, the relationships
	\begin{align}
		\mathbf{e}^{(11)'}_{(\mu \nu)} + \mathbf{e}^{(12)'}_{(\mu \nu )} &= \mathbf{e}^{(11)}_{(\mu \nu)} + \mathbf{e}^{(12)}_{(\mu \nu )}&
		\mathbf{e}^{(22)'}_{(\mu \nu)} + \mathbf{e}^{(12)'}_{(\mu \nu )} &= \mathbf{e}^{(11)}_{(\mu \nu)} + \mathbf{e}^{(12)}_{(\mu \nu )}\\
		\mathbf{e}^{(11)\alpha'}_{\alpha} + \mathbf{e}^{(12)\alpha'}_{\alpha} &= \mathbf{e}^{(11)\alpha}_{\alpha} + \mathbf{e}^{(12)\alpha}_{\alpha}&
		\mathbf{e}^{(22)\alpha'}_{\alpha} + \mathbf{e}^{(12)\alpha'}_{\alpha} &= \mathbf{e}^{(11)}_{(\mu \nu)} + \mathbf{e}^{(12)}_{(\mu \nu )}
	\end{align}
	are obtained under constraints,
	\begin{align}
		\mathbf{e}_{(11)}\Omega_{11}^{-1} + \mathbf{e}_{(12)}\Omega_{21}^{-1} &=0;&
		\mathbf{e}_{(22)}\Omega_{11}^{-1} + \mathbf{e}_{(12)}\Omega_{11}^{-1} &=0
	\end{align}
	Considering individually, it yields.
	\begin{align}
		\mathbf{e}^{(11)\alpha}_{\alpha} &= \mathbf{e}_{(11)}A_\alpha A^\alpha\\\label{Scalar phontos AA}
		\mathbf{e}^{(22)\alpha}_{\alpha} &= \mathbf{e}_{(22)}U_\alpha U^\alpha\\ 
		\mathbf{e}^{(+-)\alpha}_{\alpha} &= \left( \mathbf{e}_{(33)} + \mathbf{e}_{(44)}\right)V_\alpha^+ V^{\alpha-} \\
		\mathbf{e}^{(+1)\alpha}_{\alpha}+ \mathbf{e}^{(-1)\alpha}_{\alpha} &= \sqrt{2}\left[\left(\mathbf{e}_{(13)} + i\mathbf{e}_{(14)}\right)A_\alpha V^{\alpha+}  + \left(\mathbf{e}_{(13)} - i\mathbf{e}_{(14)}\right)A_\alpha V^{\alpha -} \right]\\
		\mathbf{e}^{(+2)\alpha}_{\alpha}+ \mathbf{e}^{(-2)\alpha}_{\alpha} &= \sqrt{2}\left[\left(\mathbf{e}_{(23)} + i\mathbf{e}_{(24)}\right)U_\alpha V^{\alpha +}  + \left(\mathbf{e}_{(23)} - i\mathbf{e}_{(24)}\right)U_\alpha V^{\alpha -} \right]\\
		\mathbf{e}^{(++)\alpha}_{\alpha} + \mathbf{e}^{(--)\alpha}_{\alpha} &= \frac{1}{2}\left(\mathbf{e}_{(33)}-\mathbf{e}_{(44)}\right)V_\alpha^+V^{\alpha+} + \frac{1}{2}\left(\mathbf{e}_{(33)}-\mathbf{e}_{(44)}\right)V_\alpha^-V^{\alpha-}\\
		\mathbf{e}^{(++34)\alpha}_{\alpha} &+ \mathbf{e}^{(--34)\alpha}_{\alpha} =  i\mathbf{e}_{(34)}V_\alpha^+V^{\alpha+} -i\mathbf{e}_{(34)}V_{\alpha}^-V^{\alpha -} \label{Scalar phontos V+V-}
	\end{align}
	transforming as
	\begin{align}
		\mathbf{e}^{(11)'}_{(\mu \nu)} &= \mathbf{e}^{(11)}_{(\mu \nu)}; &
		\mathbf{e}^{(22)'}_{(\mu \nu)} &= \mathbf{e}^{(11)}_{(\mu \nu)};&
		\mathbf{e}^{(12)'}_{(\mu \nu)} &= \mathbf{e}^{(12)}_{(\mu \nu)}\\
		\mathbf{e}^{(11)\alpha'}_{\alpha} &= \mathbf{e}^{(11)\alpha}_{\alpha};&
		\mathbf{e}^{(22)\alpha'}_{\alpha} &= \mathbf{e}^{(11)\alpha}_{\alpha};&
		\mathbf{e}^{(12)\alpha'}_{\alpha} &= \mathbf{e}^{(12)\alpha}_{\alpha}	
	\end{align}
	under conditions
	\begin{align}
		\mathbf{e}_{(11)}\Omega_{11}^{-1} &= 0 ;& \mathbf{e}_{(22)}\Omega_{21}^{-1} &= 0&
		\mathbf{e}_{(12)}\Omega_{11}^{-1} &= 0 ;& \mathbf{e}_{(12)}\Omega_{21}^{-1} &= 0
	\end{align}
	
	For charged symmetric collective fields, one gets
	\begin{align}
		\mathbf{e}^{(+-)'}_{(\mu \nu)} &=   \mathbf{e}^{(+-)}_{(\mu \nu)}& \mathbf{e}^{(+-)\alpha'}_{\alpha} &=   \mathbf{e}^{(+-)\alpha}_{\alpha}
	\end{align}
	under the following the coefficients relationships.
	\begin{align}
		\mathbf{e}_{(33)}\Omega_{31}^{-1} &= 0 ;& \mathbf{e}_{(44)}\Omega_{31}^{-1} &= 0;&
		\mathbf{e}_{(33)}\Omega_{41}^{-1} &= 0 ;& \mathbf{e}_{(44)}\Omega_{41}^{-1} &= 0
	\end{align}
	
	Other relationships are obtained
	\begin{align}
		\mathbf{e}^{(+1)'}_{(\mu \nu)} + \mathbf{e}^{(-1)'}_{(\mu \nu)} &= 	e^{iq \alpha}\mathbf{e}^{(+1)}_{(\mu \nu)} + e^{-iq \alpha}\mathbf{e}^{(-1)}_{(\mu \nu)}\\ 
		\mathbf{e}^{(+2)'}_{(\mu \nu)} + \mathbf{e}^{(-2)'}_{(\mu \nu)} &= 	e^{iq \alpha}\mathbf{e}^{(+2)}_{(\mu \nu)} + e^{-iq \alpha}\mathbf{e}^{(-2)}_{(\mu \nu)}\\
		\mathbf{e}^{(++)'}_{(\mu \nu)} + \mathbf{e}^{(--)'}_{(\mu \nu)} &= e^{2iq \alpha}\mathbf{e}^{(++)}_{(\mu \nu)} + e^{-2iq \alpha}\mathbf{e}^{(--)}_{(\mu \nu)}\\
		\mathbf{e}^{(+1)\alpha'}_{\alpha} + \mathbf{e}^{(-1)\alpha'}_{\alpha} &= e^{iq \alpha}\mathbf{e}^{(+1)\alpha}_{\alpha}+ e^{-iq \alpha}\mathbf{e}^{(-1)\alpha}_{\alpha}\\
		\mathbf{e}^{(+2)\alpha'}_{\alpha}+ \mathbf{e}^{(-2)\alpha'}_{\alpha} &= e^{iq \alpha}\mathbf{e}^{(+2)\alpha}_{\alpha}+ e^{-iq \alpha}\mathbf{e}^{(-2)\alpha}_{\alpha}\\
		\mathbf{e}^{(++)\alpha'}_{\alpha} + \mathbf{e}^{(--)\alpha'}_{\alpha} &= e^{2iq \alpha}\mathbf{e}^{(++)\alpha}_{\alpha} + e^{-2iq \alpha}\mathbf{e}^{(--)\alpha}_{\alpha}
	\end{align}
	under the relationships.
	\begin{align}[h]
		\mathbf{e}_{(13)}\Omega_{31}^{-1} &= 0; & \mathbf{e}_{(14)}\Omega_{31}^{-1} &= 0;&
		\mathbf{e}_{(13)}\Omega_{41}^{-1} &= 0; & \mathbf{e}_{(14)}\Omega_{41}^{-1} &= 0\\
		\mathbf{e}_{(13)}\Omega_{11}^{-1} &= 0; & \mathbf{e}_{(14)}\Omega_{11}^{-1} &= 0;&
		\mathbf{e}_{(23)}\Omega_{31}^{-1} &= 0; & \mathbf{e}_{(24)}\Omega_{31}^{-1} &= 0;\\
		\mathbf{e}_{(23)}\Omega_{41}^{-1} &= 0; & \mathbf{e}_{(24)}\Omega_{41}^{-1} &= 0;&
		\mathbf{e}_{(23)}\Omega_{11}^{-1} &= 0; & \mathbf{e}_{(24)}\Omega_{11}^{-1} &= 0
	\end{align}
	
	and
	\begin{align}
		\mathbf{e}^{(++34)'}_{(\mu \nu)} + \mathbf{e}^{(--34)'}_{(\mu \nu)} &= e^{2iq \alpha}\mathbf{e}^{(++34)}_{(\mu \nu)} + e^{-2iq \alpha}\mathbf{e}^{(--34)}_{(\mu \nu)}\\
		\mathbf{e}^{(++34)\alpha'}_{\alpha} + \mathbf{e}^{(--34)\alpha'}_{\alpha} &= e^{2iq \alpha}\mathbf{e}^{(++34)\alpha}_{\alpha} + e^{-2iq \alpha}\mathbf{e}^{(--34)\alpha}_{\alpha}
	\end{align}
	for
	\begin{align}
		\mathbf{e}_{(34)}\Omega_{31}^{-1} &=0 ;& \mathbf{e}_{(34)}\Omega_{41}^{-1} =0 
	\end{align}
	
	Consequently, the number of constraints is less than the number of free coefficients.
	\section{Appendix Noether theorem second vision}
	
	Considering eq. (\ref{Noether charge}) and Euler-Lagrange equations, eq. (\ref{k equation}) can be rewritten. It yields the following conservation law.

	\begin{equation}\label{Noether second vision}
		\bar{K}^{\mu} + J^{\mu}_N=0
	\end{equation}
	
	Where
	\begin{equation}
		\bar{K}^{\mu} = k_1\frac{\partial L}{\partial A_{\mu}} + k_2\frac{\partial L}{\partial U_{\mu}} + k_+\frac{\partial L}{\partial V_{\mu}^+} + k_-\frac{\partial L}{\partial V_{\mu}^-}
	\end{equation}
	and
	\begin{eqnarray}
		&&\bar{K}^\mu = k_1\{ 2\mathbf{e}_{(11)}\left( \beta_2 S^{\mu \nu 2} + \rho_1 g^{\mu \nu}S_\alpha^{\alpha 1} + \rho_2 g^{\mu \nu}S_\alpha^{\alpha 2} \right)A_\nu +\nonumber 
		\\
		&& +2\mathbf{e}_{(11)}\big[ 16\rho_1 S_\alpha^{\alpha 1}+ \left(\beta_2 + 16\rho_2 S_\alpha^{\alpha 2}\right)\big] A^\mu+2\mathbf{e}_{(12)}\big(\beta_2 S^{\mu \nu 2} \nonumber
		\\
		&&+ \rho_1g^{\mu \nu}S_\alpha^{\alpha 1} + \rho_2g^{\mu \nu}S_\alpha^{\alpha 2} \big)U_\nu + 4\big(\mathbf{e}_{(11)}\mathbf{e}^{(11)(\mu \nu)} + 2g^{\mu \nu}\mathbf{e}^{(11)\alpha}_\alpha A^{\mu} \nonumber
		\\
		&&+ 2\mathbf{e}^{(+-)(\mu \nu)} + \mathbf{e}^{(12)(\mu \nu)} + g^{\mu \nu}\mathbf{e}^{(12)\alpha}_\alpha\big)A_\nu+ 8 \mathbf{e}_{(12)}\mathbf{e}^{(12)(\mu \nu)}A_\nu  \nonumber
		\\
		&& + 4\mathbf{e}_{(12)}\mathbf{e}^{(12)\alpha}_\alpha A^\mu + 4\mathbf{e}_{(12)}\mathbf{e}^{(22)(\mu \nu)}U_\nu +4\mathbf{e}_{(12)}\mathbf{e}^{(12)\alpha}_\alpha U^\mu+  \nonumber
		\\
		&&+68\mathbf{e}_{(11)}\mathbf{e}^{(22)(\mu \nu)}A_\nu + \mathbf{e}_{(11)}\mathbf{e}^{(+-)\alpha}_\alpha A^\mu + 84\mathbf{e}_{(12)}\mathbf{e}^{(22)(\mu \nu)}U_\nu \nonumber
		\\
		&&+ 72\mathbf{e}_{(12)}\mathbf{e}^{(22)\alpha}_\alpha U^\mu+72\mathbf{e}_{(12)}\mathbf{e}^{(+-)\alpha}_\alpha U^\mu\} +k_2\{ 2\mathbf{e}_{(22)}\big( \beta_1 S^{\mu \nu 1} \nonumber
		\\
		&&+ \rho_1 g^{\mu \nu}S_\alpha^{\alpha 1} + \rho_2 g^{\mu \nu}S_\alpha^{\alpha 2} \big)U_\nu +2\mathbf{e}_{(22)}\big[ \left( \beta_1 + 16\rho_1\right) S_\alpha^{\alpha 1} 
		\\
		&&+ 16\rho_2 S_\alpha^{\alpha 2} \big] U^\mu + 2\mathbf{e}_{(12)}\big(\beta_1 S^{\mu \nu 1} + \rho_1g^{\mu \nu}S_\alpha^{\alpha 1} + \rho_2g^{\mu \nu}S_\alpha^{\alpha 2} \big)A_\nu\nonumber
		\\
		&&+ 8\mathbf{e}_{(22)}\mathbf{e}^{(22)\alpha}_\alpha U^\mu + 8\mathbf{e}_{(12)}\mathbf{e}^{(12)(\mu \nu)}_\alpha U_\nu + 8\mathbf{e}_{(22)}\mathbf{e}^{(+-)(\mu \nu)}U_\nu +\nonumber
		\\
		&&+ 68\mathbf{e}_{(22)}\mathbf{e}^{(12)(\mu \nu)}U_\nu + 8\mathbf{e}_{(12)}\mathbf{e}^{(22)(\mu \nu)}A_\nu 
		+4\mathbf{e}_{(12)}\mathbf{e}^{(12)\alpha}_\alpha A^\mu+\nonumber
		\\
		&&+ 68\mathbf{e}_{(12)}\mathbf{e}^{(12)\alpha}_\alpha U^\mu + \mathbf{e}_{(22)}\mathbf{e}^{(+-)\alpha}_\alpha U^\mu +8\mathbf{e}_{(12)}\mathbf{e}^{(11)(\mu \nu)}A_\nu \nonumber
		\\
		&&+ 72\mathbf{e}_{(12)}\mathbf{e}^{(11)\alpha}_\alpha A^\mu + 72\mathbf{e}_{(12)}\mathbf{e}^{(+-)\alpha}_\alpha A^\mu \}\nonumber
	\end{eqnarray}
	
	Differently from eq. (\ref{equação da simetria}), eq. (\ref{Noether second vision}) does not produce an equation of motion but just a constraint.
\end{appendices}

\end{document}